\documentclass[11pt,a4paper]{article}
\usepackage{amssymb}
\author{Emmanuel
Kohlprath\footnote{e-mail: emmanuel.kohlprath@cpht.polytechnique.fr} \\
\\
\textit{Centre de Physique Th{\'e}orique}\\ \textit{{\'E}cole Polytechnique}\\
\textit{F-91128 Palaiseau-Cedex}\\ \textit{France}}
\title{\vspace*{-3cm}\begin{flushright}
{\small CPHT-RR-109.1103}\end{flushright}
\vspace*{2cm} Induced gravity in $\mathbb{Z}_N$
orientifold models\footnote{Research partialy founded by the EEC under
contracts HPRN-CT-2000-00122, HPRN-CT-2000-00131 and
HPRN-CT-2000-00148.}} 

\newcommand{\be}{\begin{equation}}
\newcommand{\ee}{\end{equation}}
\newcommand{\ba}{\begin{eqnarray}}
\newcommand{\ea}{\end{eqnarray}}
\newcommand{\no}{\nonumber}

\date{}

\begin{document}
\maketitle

\begin{abstract} 
We consider non-compact $\mathbb{Z}_N$ orientifold models of type IIB
super\-string theory with four-dimensional gravity induced on a set of
coincident D3-branes. For the models with odd $N$ the contribution to
the one-loop renormalization of the Planck mass is shown to come only
from the torus and to be $O(N)$ as the contributions from
annulus, Moebius strip and Klein bottle cancel. One can therefore
realize the Dvali-Gabadadze-Porrati idea that four-dimensional gravity
is induced by quantum effects at the one-loop level by considering
large $N$.
\end{abstract}

\section{Introduction}

Today we know of several ways to realize four-dimensional gravity 
(see e.g. \cite{Kiritsisreview1} for a review). First, we get a
four-dimensional Einstein-Hilbert term from 
a D-dimensional one if we compactify $D-4$ dimensions. Second, we can
have a cosmological constant in the bulk and on a codimension one
3-brane. This is the Randall-Sundrum mechanism
\cite{RandallSundrum99a} and \cite{RandallSundrum99b}. Third, we can
have brane induced gravity, i.e. we can have a four-dimensional
Einstein-Hilbert term that is induced on a 3-brane by the
fields that live on the 3-brane. This has been put forward by Dvali,
Gabadadze and Porrati (\cite{DvaliGabadadzePorrati00a} --
\cite{DvaliGabadadze00}). Obviously we can also combine these three
ideas in a given model.\\

Both Randall-Sundrum and
Dvali-Gabadadze-Porrati originally work in a five-dimensional bulk
(i.e. with codimension one). Whereas the Randall-Sundrum setup leads to 
a four-dimensional behaviour of gravity at long distances and a
five-dimensional behaviour of gravity at short distances (that is
similar to compactification) in the case of the
Dvali-Gabadadze-Porrati setup gravity is four-dimensional at short
distances and five-dimensional at long distances. Therefore to be
consistent with experiments and astronomical observations the
cross-over scale has either to be astronomically large or one has
to compactify the extra dimension
\cite{DvaliGabadadzeKolanovicNitti01}. 
In the later case Kaluza-Klein graviton emission is suppressed in the
ultraviolet and therefore also the energy loss from brane to
bulk. Constraints from experiments on the size of the compact
dimension are therefore less stringent than for standard
compactifications.\\

With a codimension greater than one the effects of a finite brane
thickness can no longer be neglected 
(\cite{KiritsisTetradisTomaras01} -- \cite{Kolanovic03a}),
but the  
conclusions on the cross-over scale or on the need for compactification 
remain the same. On the other hand one can also merge the setups of
Randall-Sundrum and of Dvali-Gabadadze-Porrati 
\cite{KiritsisTetradisTomaras02}.\\

In a complete theory we need four-dimensional gravity and we will
consider how this can be achieved in superstring theory. 
We may ask the question if there are string
models with brane induced gravity where the one-loop induced
four-dimensional Planck mass is large in string units. For heterotic
string theory such corrections vanish for $\mathcal{N}\geq1$
supersymmetry (\cite{KiritsisKounnas95} --
\cite{Kiritsisreview}). For type II vacuua such corrections can be
non-vanishing for $\mathcal{N}\leq 2$ supersymmetry
\cite{KiritsisKounnas95},\cite{AntoniadisFerrareMinasianNarain97}. 
In particular for
a background of the type $M_4\times CY_3$, where $M_4$ is
four-dimensional Minkowski space and $CY_3$ is a Calabi-Yau, the
one-loop correction is proportional to the Euler number 
(\cite{AntoniadisFerrareMinasianNarain97} -- \cite{Antoniadis02}).\\

In this paper we will consider type I/orientifold vacuua. These models
can have gauge and matter fields on the D3-branes
that come close to the standard model (and supersymmetric
generalizations thereof). Explicitely we will consider
models that are non-compact (non-standard) orientifolds of type IIB
superstring theory on symmetric $\mathbb{Z}_N$ orbifolds and we will
compute the induced gravity on a set of coincident D3-branes. 
The reasons are the following: As in this setup we have the gauge
fields, matter fields and gravity localized on the D3-branes we do not
need to compactify. Besides being interesting for its own, this has
the advantage that we can have arbitrary $N$ and may consider the
large $N$ limit and that we can have an arbitrary number of D3-branes
(no untwisted tadpole cancellation condition).\\

Our orientifold models are similar to the ones discussed in
\cite{Kakushadze98} and \cite{Kakushadze01} but more general as we consider
models with arbitrary possibly large $N$. In
\cite{IbanezRabadanUranga98} the anomaly cancellation in these models
has been analysed. We will determine the renomalization of the
four-dimensional Planck mass. Whereas in \cite{Kakushadze98}
and \cite{Kakushadze01} the contributions to the Planck mass are only
stated to exist and to be determined by the string scale we
explicitely determine them by a string computation and show that they
are large if $N$ is large. This is a
generalization of the computation of \cite{Antoniadis96} and
\cite{Antoniadis02} that considered compacification on K3. We can then 
compare the torus versus the annulus, Moebius strip and Klein bottle
contributions. \\

Writing the one-loop renormalization of the four-dimensional Planck
mass as
\be
\Delta{\mathcal
L}_{\textrm{eff}}^{\textrm{1-loop}}=\delta M_s^2 \sqrt{-g}R
\ee
we will show that the torus contribution is $O(N)$ and that annulus,
Moebius strip and Klein bottle contributions cancel. Therefore, by
considering large $N$ the one-loop contribution can be arbitrary
large. The number of gauge and matter fields on the D3-branes is not
growing with $N$ and we can in principle find models that are quite
close to supersymmetric generalizations of the standard model. \\

There is also a different string theory realization of induced gravity
presented in \cite{KiritsisTetradisTomaras01} that is based on 
the orientifold of K3 and two extra compactified dimensions as in
\cite{Antoniadis96}. There the contibution to the four-dimensional
Planck mass comes only from the Kaluza-Klein tower from annulus,
Moebius strip and Klein bottle as the torus does not contribute.\\ 

In section \ref{sectionorbifold} we consider $\mathbb{Z}_N$ orbifolds
of type IIB and review the contribution to the Planck mass from the
torus. In sections \ref{sectionorientifold} we consider $\mathbb{Z}_N$
orientifolds of type IIB and compute the contribution to the Planck
mass from the annulus, the Moebius strip and the Klein bottle. In section
\ref{sectionconclusions} we give our conclusions. Some details of the
computations are left to appendices.

\section{$\mathbb{Z}_N$ orbifolds} \label{sectionorbifold}

In order to have localized twisted sectors and therefore a localized
Einstein term and in order to have a parameter $N$ that we may e.g.
assume to be large we consider $\mathbb{Z}_N$ orbifolds of type IIB
superstring theory in this section. In the compact case we compactify on
$M^4\times T^6/{\mathbb Z}_N$, where $M^4$ is four-dimensional Minkowski
space. In the non-compact case the background is $M^4\times
\mathbb{R}^6/{\mathbb Z}_N$.  We have ${\mathcal N}=2$ supersymmetry
in $d=4$. In section \ref{sectiono1} we will first review how the
contribution of the torus to the one-loop renormalization of the
Planck mass is determined by the Euler number or the second helicity
supertrace. In section \ref{sectiono2} we then show how the
second helicity supertrace follows from the helicity generation
partition function. In section \ref{sectiono3} we analyse the
large $N$ limit. In \ref{sectiono4} we compute the torus contribution
from a two graviton amplitude in order to fix the vertex operator
normalization that we will need in section \ref{sectionorientifold}.

\subsection{The second helicity supertrace} \label{sectiono1}

The helicity supertraces are defined by (see \cite{Kiritsisreview})
\be
B_{2n}=\textrm{Str}\left[\lambda^{2n}\right]\qquad
n\in{\mathbb N},
\ee
where the $\lambda$'s are the helicity eigenvalues.
The contribution of the ${\mathcal N}=2$ supergravity multiplet and
of ${\mathcal N}=2$ vector multiplets to the second helicity
supertrace $B_2$ is $1$, whereas the contribution of ${\mathcal N}=2$
hyper multiplets is $-1$. This gives
\be
B_2=1+n_V-n_H,
\ee
where $n_V$ and $n_H$ count the number of vector and hyper
multiplets. The ${\mathbb Z}_N$ orbifolds we consider are
singular limits of Calabi-Yau 3-folds with hodge numbers
\be
h^{1,1}=n_V,\qquad h^{2,1}=n_H-1,
\ee
where we have subtracted the universal hyper multiplet. The Euler
number is
\be
\chi=2\left(h^{1,1}-h^{2,1}\right).
\ee
This gives
\be
B_2=\frac{1}{2}\chi.
\ee
The only one-loop surface is the torus $\mathcal{T}$.
In \cite{AntoniadisFerrareMinasianNarain97} it was shown (see also
\cite{Antoniadis02} and \cite{Antoniadis03}) that this
gives a one-loop renormalization of the Planck mass of\footnote{We
have $M_s^2=1/\alpha'$.}
\be
\Delta{\mathcal L}_{\textrm{eff}}^{\textrm{1-loop}}
=\delta_{\mathcal{T}} M_s^2 \sqrt{-g}R=\frac{1}{12\pi}\chi M_s^2
\sqrt{-g}R=\frac{1}{6\pi}B_2 M_s^2 \sqrt{-g}R.
\label{Leffoneloop}
\ee
With $\lambda=\lambda_L+\lambda_R$ we get from the helicity generating
partition function $Z(v,\bar v)=\textrm{str}(q^{L_0}\bar q^{\bar
L_0}\exp(2\pi i(v\lambda_R-\bar v\lambda_L)))$ the second helicity
supertrace as (see \cite{Kiritsisreview}) 
\be
B_2=-\left.\Bigl(\frac{1}{2\pi i}\partial_v-\frac{1}{2\pi i}\partial_{\bar
v}\Bigr)^2 Z(v,\bar v)\right|_{v=\bar v=0}.\label{B2partition}
\ee

\subsection{The helicity generating partition function}\label{sectiono2}

Let us define the complex bosons as $Z^i=X^{2i+2}+iX^{2i+3}$, $i=1,2,3$.
The helicity generating torus partition function for the ${\mathbb
Z}_N$ orbifold of type IIB is\footnote{In the non-compact case the
partition function for the (0,0) sector is normalized with respect to
the ten-dimensional volume and the partition function for the
remaining sectors with respect to the four-dimensional volume. To
derive the spectrum and the helicity supertraces one starts with the
unintegrated partition function.} (see also \cite{Kohlprath02})
\ba
&&\hspace*{-2cm}Z^{(0,0)}(v,\bar v)=N_0(N)\int\limits_{\mathcal
F}\!\frac{d^2\tau}{\tau_2^2}Z_X^2(\tau)
\left.\left[Z_\psi^+(v,\tau)Z_\psi^+(v,\tau)^*\right]\right|_{h=g=0}
\prod_{i=1}^3 Z_i\left[\hspace*{-6pt}\begin{array}{c}0\\0
\end{array}\hspace*{-6pt}\right](\tau) \\
&&\hspace*{-2cm}Z'(v,\bar v)=N_0(N)\int\limits_{\mathcal
F}\!\frac{d^2\tau}{\tau_2^2}Z_X^2(\tau)
\hspace*{-1cm}\sum_{\begin{array}{c}
h,g=0\\(h,g)\not=(0,0)\end{array}}^{N-1}\hspace*{-1cm}
Z_\psi^+(v,\tau)Z_\psi^+(v,\tau)^*
\prod_{i=1}^3 Z_i\left[\hspace*{-6pt}\begin{array}{c}hv_i\\gv_i
\end{array}\hspace*{-6pt}\right](\tau),\label{ZNZMpartitionfunction}
\ea
where 
\ba
Z_X^2(\tau)&=&\frac{1}{\tau_2}\frac{1}{|\eta(\tau)|^4}
\label{ZNZMpartitionfunction1}\\
Z_i\left[\hspace*{-6pt}\begin{array}{c}0\\0\end{array}\hspace*{-6pt}\right]
(\tau)&=&\frac{\Gamma_{2,2}}{|\eta(\tau)|^4}\ 
(\Gamma_{2,2}\textrm{ is the }(2,2)\textrm{ lattice sum})
\label{ZNZMpartitionfunction2}\\
Z_i\left[\hspace*{-6pt}\begin{array}{c}hv_i\\gv_i\end{array}
\hspace*{-6pt}\right]
(\tau)&=&C^{(N)}(hv_i,gv_i)\left|\frac{\eta(\tau)}{\theta\left[\hspace*{-6pt}
\begin{array}{c}1/2+hv_i\\1/2+gv_i\end{array}\hspace*{-6pt}\right](0,\tau)}
\right|^2\textrm{ for }(hv_i,gv_i)\not=(0,0)\no\\&&\\
Z_\psi^+(v,\tau)&=&\frac{\xi(v)}{2}\frac{1}{\eta(\tau)^4}
\sum_{\alpha,\beta=0}^{1} (-)^{\alpha+\beta+\alpha\beta}
\theta\left[\hspace*{-6pt}\begin{array}{c}\alpha/2\\
\beta/2\end{array}\hspace*{-6pt}\right](v,\tau)
\theta\left[\hspace*{-6pt}\begin{array}{c}\alpha/2+hv_1\\\beta/2
+gv_1\end{array}\hspace*{-6pt}\right](0,\tau)\no\\&&\times 
\theta\left[\hspace*{-6pt}\begin{array}{c}\alpha/2+hv_2\\\beta/2
+gv_2\end{array}\hspace*{-6pt}\right](0,\tau)
\theta\left[\hspace*{-6pt}\begin{array}{c}\alpha/2+hv_3\\\beta/2
+gv_3\end{array}\hspace*{-6pt}\right](0,\tau)\\
\xi(v)&=&\frac{\sin\pi v}{\pi}\frac{\left.\partial_u
\theta\left[\hspace*{-6pt}\begin{array}{c}1/2\\
1/2\end{array}\hspace*{-6pt}\right](u,\tau)\right|_{u=0}}
{\theta\left[\hspace*{-6pt}\begin{array}{c}1/2\\
1/2\end{array}\hspace*{-6pt}\right](v,\tau)}.
\ea
For the untwisted sector in the compact case
\be
\left|C^{(N)}(0,gv_i)\right|=4\left(\sin(\pi gv_i)\right)^2
\ee
whereas in the non-compact case $\left|C^{(N)}(0,gv_i)\right|=1$ as
there is an extra factor of $1/(4\left(\sin(\pi gv_i)\right)^2)$
comming from the integration over non-compact momenta\footnote{The
author thanks E. Dudas and P. Vanhove for clarifying this point}.
For the twisted sectors ($h\not=0$)
$\left|C^{(N)}(hv_i,gv_i)\right|$ counts the fixed points
multiplicity that is always 1 in the non-compact case.
The torus partition function
\be
Z_{\mathcal{T}}=Z(0,0)
\ee
is as expected modular invariant (use $\xi(0)=1$).
Using the Riemann identity we get
\ba
Z_\psi^+(v,\tau)&=&\frac{\xi(v)}{\eta(\tau)^4}
\theta\left[\hspace*{-6pt}\begin{array}{c}1/2\\
1/2\end{array}\hspace*{-6pt}\right](v/2,\tau)
\theta\left[\hspace*{-6pt}\begin{array}{c}1/2-hv_1\\
1/2-gv_1\end{array}\hspace*{-6pt}\right](v/2,\tau)\no\\&&\times
\theta\left[\hspace*{-6pt}\begin{array}{c}1/2-hv_2\\
1/2-gv_2\end{array}\hspace*{-6pt}\right](v/2,\tau)
\theta\left[\hspace*{-6pt}\begin{array}{c}1/2-hv_3\\
1/2-gv_3\end{array}\hspace*{-6pt}\right](v/2,\tau).
\ea
Let ${\mathcal M}$ be the set of elements $\{(h,g)\}$ that solve
\ba
hv_1=gv_1&=&0\qquad {\rm mod 1}\\
{\rm or}\qquad hv_2=gv_2&=&0\qquad {\rm mod 1}\\
{\rm or}\qquad hv_3=gv_3&=&0\qquad {\rm mod 1}.
\ea
Obviously $(0,0)\in{\mathcal M}$.
From (\ref{B2partition}) we get the second helicity supertrace
for the ${\mathbb Z}_N$ orbifold
\be
B_2=\frac{1}{2}N_0(N)\sum\limits_{\begin{array}{c}h,g=0\\(h,g)\not\in{\mathcal
M}\end{array}}^{N-1}
\left|C^{(N)}(hv_1,gv_1)\right.
\left.C^{(N)}(hv_2,gv_2) C^{(N)}(hv_3,gv_3)\right|.\no\\
\label{formulaforB2}
\ee
Notice that in the non-compact case only the twisted states
($h\not=0$) contribute as they are the ones that are localized in
four-dimensions and induce the four-dimensional Planck mass.
The normalization
\be
N_0(N)=\frac{1}{N}\label{partitionfunctionnormalization}
\ee
is fixed by matching the massless spectrum
that one gets from the operator approach with the one one derives from 
the helicity generating partition function. In appendix
\ref{Z3example} we first show this for the example of the
$\mathbb{Z}_3$ orbifold and then give the proof for prime $N$. The
proof for the case with general $N\in \mathbb{N}$ is only scatched as
it is straight forward and lenghty.

\subsection{Large $N$ behaviour of $\mathbb{Z}_N$ orbifolds}\label{sectiono3}
\label{largeNbehaviour}

For a non-compact ${\mathbb Z}_N$ orbifold the second helicity
supertrace comming from twisted sectors is $B_2^T=n_V^T-n_H^T$. We
have
\ba
N \textrm{\ even:\ }&& n_V^T=\frac{N-2}{2}+1=\frac{N}{2}, \quad
n_H^T=0\label{generaltwistedspectrum1}\\
N \textrm{\ odd:\ }&& n_V^T=\frac{N-1}{2}, \quad
n_H^T=0.\label{generaltwistedspectrum2}
\ea
and therefore we have the behavior
\be
B_2^T\stackrel{N\rightarrow\infty}{\longrightarrow}\frac{N}{2}+O(1)
\ee
that gives using (\ref{Leffoneloop})
\be
\Delta{\mathcal L}_{\textrm{eff}}^{\textrm{1-loop}}
\stackrel{N\rightarrow\infty}{\longrightarrow}\frac{N+O(1)}{12\pi} M_s^2
\sqrt{-g}R.\label{largeNtoruscontribution}
\ee
In the compact case the shift vector $v$ for a $\mathbb{Z}_N$ orbifold 
has to be such that the orbifold acts crystallographically. In the
non-compact case this is obviously not necessary. For the
$\mathbb{Z}_N$ orbifold with odd $N$ we can e.g. choose the shift
vector $v=\left(\frac{1}{N},\frac{1}{N},-\frac{2}{N}\right)$. Then
${\mathcal M}=\{(0,0)\}$ and (\ref{formulaforB2}) gives the twisted
contribution (\ref{generaltwistedspectrum2}).

\subsection{The Planck mass from the two graviton amplitude}
\label{sectiono4}

In this section we compute the torus contribution
from a two graviton amplitude in order to fix the vertex operator
normalization that we will need in section \ref{sectionorientifold}.

\subsubsection{Matching amplitudes and effective actions}

Let us define
\ba
g_{\mu\nu}&=&\eta_{\mu\nu}+\kappa h_{\mu\nu}\\
\Gamma^{\mu}_{\ \nu\rho}&=&\frac{1}{2}g^{\mu\lambda}\left(\partial_\rho g_{\lambda\nu} - \partial_\lambda g_{\nu\rho} + \partial_\nu g_{\rho\lambda}\right)\\
R^{\mu}_{\ \nu\rho\sigma}&=& \partial_\rho\Gamma^{\mu}_{\ \nu\sigma} - \partial_\sigma\Gamma^{\mu}_{\ \nu\rho} + \Gamma^{\mu}_{\ \rho\lambda}\Gamma^{\lambda}_{\ \nu\sigma} -  \Gamma^{\mu}_{\ \sigma\lambda}\Gamma^{\lambda}_{\ \nu\rho}. 
\ea
With the graviton
\be
h_{\mu\nu}(x)=\int\!\frac{d^4p}{(2\pi)^4}\ e^{ipx}\varepsilon_{\mu\nu},
\ee
neglecting terms proportional to $p_1^2, p_2^2, p_1\cdot p_2,
p_{1\mu} \varepsilon^{\mu\nu}_1, p_{2\rho}\varepsilon^{\rho\sigma}_2,
\eta_{\mu\nu}\varepsilon^{\mu\nu}_1,
\eta_{\rho\sigma}\varepsilon^{\rho\sigma}_2$
(due to the fact that gravitons are massless and that the polarizarion
tensors are physical) and keeping the momenta arbitrary (no momentum
conservation) we find in momentum space
\be
\left.\sqrt{-g}R\right|_{O(\kappa^2)}\rightarrow-\frac{\kappa^2}{8} \Bigl( \eta_{\mu\rho}p_{1\sigma}p_{2\nu} + \eta_{\mu\sigma}p_{1\rho}p_{2\nu} +  \eta_{\nu\rho}p_{1\sigma}p_{2\mu} + \eta_{\nu\sigma}p_{1\rho}p_{2\mu}\Bigr) \varepsilon^{\mu\nu}_1\varepsilon^{\rho\sigma}_2.
\ee
It is enough to consider only one tensor structure as it follows from
covariance that the only term in the effective action that contributes
in second order in momentum is $\sqrt{-g}R$. Let us write the (off-shell) two graviton amplitude as
\be
\left.A^{(2)}\right|_{O(p^2)}=-\frac{1}{4C_m}\delta\Bigl( \eta_{\mu\rho}p_{1\sigma}p_{2\nu} + \eta_{\mu\sigma}p_{1\rho}p_{2\nu} +  \eta_{\nu\rho}p_{1\sigma}p_{2\mu} + \eta_{\nu\sigma}p_{1\rho}p_{2\mu}\Bigr) \varepsilon^{\mu\nu}_1\varepsilon^{\rho\sigma}_2,\label{definingdelta}
\ee
where the momenta are measured in string units (i.e. they are
dimensionless) and we have introduces a matching coefficient $C_m$
that we will determine and that accommodates the fact that we use
vertex operators that are not normalized properly.
Then the contribution to the effective action is precisely
\be
\Delta\mathcal{L}_{\textrm{eff}}=M_s^2\delta \sqrt{-g}R,
\ee
where a factor of $\frac{1}{2}$ from Bose symmetry is taken into
account.

\subsubsection{The two graviton amplitude}

The Einstein term is CP even and gets contributions from the even-even
and the odd-odd spin structure two graviton amplitudes. Using the
notation of appendix \ref{appendixamplitudedetails}
the even-even spin structure two graviton amplitude is
\ba
A^{(2)}_{(e-e)}&=&\sum_{(\alpha,\beta)=0,1}^{even}
\sum_{(\bar\alpha,\bar\beta)=0,1}^{even}
\int\limits_{\Gamma}\!\frac{d^2\tau}{\tau_2^2}
(-)^{\alpha+\beta+\alpha\beta} (-)^{\bar\alpha+\bar\beta+\bar\alpha\bar\beta}  Z(\tau,\bar\tau, (\alpha,\beta), (\bar\alpha,\bar\beta))\no\\&&\times \int\!d^2 z_{1}\int\!d^2 z_{2} \langle V^{(0,0)}(z_1,\bar z_1) V^{(0,0)}(z_2,\bar z_2) \rangle_{ (\alpha,\beta),(\bar\alpha,\bar\beta)}
\ea
with the graviton vertex operator in the $(0,0)$-ghost picture
\be
V^{(0,0)}(z,\bar z)=-\frac{2g_s}{\alpha'}\, \varepsilon_{\mu\nu}\, :\,
\left(i\partial X^\mu-\frac{\alpha'}{2}\psi^\mu p\cdot\psi\right)
\left(i\bar\partial X^\nu+\frac{\alpha'}{2}\tilde\psi^\nu
p\cdot\tilde\psi\right) e^{ip\cdot X}:\, .
\label{00picturegravitonvertexoperator}
\ee
The piece in second order in momentum  vanishes due to
(\ref{bosonicG2}) and (\ref{bosonicG3}) and there are no pinching
contributions from $O(p^4)$. The other possible
contribution comes from the odd-odd spin structure two
graviton amplitude
\ba
A^{(2)}_{(o-o)}&=&\int\limits_{\Gamma}\!\frac{d^2\tau}{\tau_2^2}
Z(\tau,\bar\tau, (1,1), (1,1))\int\!\!d^2 z_{1}\int\!\!d^2 z_{2}
\no\\&&\times
\langle V^{(0,0)}(z_1,\bar z_1) V^{(-1,-1)}(z_2,\bar z_2)
X^{pc}(z_{pc},\bar z_{pc}) \rangle,
\ea
where the $(-1,-1)$-ghost picture vertex operator is
\be
V^{(-1,-1)}=g_s\varepsilon_{\mu\nu}:\psi^\mu\tilde\psi^\nu e^{ip\cdot
X}:
\ee
and the picture changing operator is
\be
X^{pc}=\partial X^\alpha\psi_\alpha\bar\partial X^\beta\tilde\psi_\beta.
\ee
Actually to get the right tensor structure we have to consider a
different distribution of the pictures. This is due to our choice of
off-shell procedure and can be avoided by considering amplitudes with
more gravitons \footnote{The author thanks
P. Vanhove for clarifying this point.}. We start with
\ba
A^{(2)}_{(o-o)}&=&\int\limits_{\Gamma}\!\frac{d^2\tau}{\tau_2^2}
Z(\tau,\bar\tau, (1,1), (1,1))\int\!\!d^2 z_{1}\int\!\!d^2 z_{2}
\no\\&&\times
\langle V^{(0,-1)}(z_1,\bar z_1) V^{(-1,0)}(z_2,\bar z_2)
X^{pc}(z_{pc},\bar z_{pc}) \rangle.
\ea
The amplitude is independent on the position of the picture changing
operator $z_{pc}$.
We have 4 fermion zero modes and the first non-vanishing correlator
has 4 fermions
\ba
&&\langle\psi^\mu(z_1)\psi^\nu(z_1)\psi^\rho(z_2)\psi^\sigma(z_{pc})\rangle=
\varepsilon^{\mu\nu\rho\sigma}\frac{1}{\alpha'}g_1(z_1,z_2,z_{pc},\tau)\no\\&& 
\langle\tilde\psi^\mu(\bar z_1)\tilde\psi^\nu(\bar
z_2)\tilde\psi^\rho(\bar z_2)\tilde\psi^\sigma(\bar z_{pc})\rangle=
\varepsilon^{\mu\nu\rho\sigma}\frac{1}{\alpha'}g_2(z_1,z_2,z_{pc},\tau)^*.
\ea
Using
\be
\varepsilon_\alpha^{\
\mu\gamma\rho}\varepsilon^{\alpha\nu\delta\sigma}=-\det\left(\begin{array}{ccc}\eta^{\mu\nu}&\eta^{\mu\delta}&\eta^{\mu\sigma}\\
\eta^{\gamma\nu}&\eta^{\gamma\delta}&\eta^{\gamma\sigma}\\
\eta^{\rho\nu}&\eta^{\rho\delta}&\eta^{\rho\sigma} \end{array}\right)
\label{epsilonsquare}
\ee
and neglecting terms proportional to $p_1^2, p_2^2, p_1\cdot p_2,
p_{1\mu} \varepsilon^{\mu\nu}_1, p_{2\rho}\varepsilon^{\rho\sigma}_2,
\eta_{\mu\nu}\varepsilon^{\mu\nu}_1,
\eta_{\rho\sigma}\varepsilon^{\rho\sigma}_2$
(what leaves only one term from (\ref{epsilonsquare})) we find the
piece in second order in momentum
\ba
\left.A^{(2)}_{(o-o)}\right|_{O(p^2)}&=&g_s^2\int\limits_{\Gamma}\!
\frac{d^2\tau}{\tau_2^2} Z(\tau,\bar\tau, (1,1), (1,1))
\frac{1}{\alpha'^2} h(\tau,\bar \tau)\no\\&&\times\frac{\alpha'}{8}
\Bigl( \eta_{\mu\rho}p_{1\sigma}p_{2\nu} +
\eta_{\mu\sigma}p_{1\rho}p_{2\nu} +  \eta_{\nu\rho}p_{1\sigma}p_{2\mu}
+ \eta_{\nu\sigma}p_{1\rho}p_{2\mu}\Bigr)
\varepsilon^{\mu\nu}_1\varepsilon^{\rho\sigma}_2,\no\\&&
\ea
where
\be
h(\tau,\bar\tau)=\int\!\! d^2z_1 \int\!\! d^2z_2\ g_1g_2^*
\langle\partial X(z_{pc},\bar z_{pc}) \bar\partial X(z_{pc},\bar
z_{pc}) \rangle.
\ee
From now on we measure positions and momenta in string units
($\frac{1}{\alpha'}d^2z\rightarrow
d^2z,\alpha'p_{\mu}p_{\nu}\rightarrow p_{\mu}p_{\nu}$).
Comparing with (\ref{definingdelta}) we get
\be
\delta_{\mathcal{T}}=-\frac{1}{2}g_s^2C_m\int\limits_{\Gamma}\!
\frac{d^2\tau}{\tau_2^2} Z(\tau,\bar\tau, (1,1), (1,1))
h(\tau,\bar\tau).
\ee
The odd-odd partition function (see (\ref{ZNZMpartitionfunction})) is
proportional to $\left|\theta\left[\hspace*{-6pt}\begin{array}{c}1/2\\
1/2\end{array}\hspace*{-6pt}\right](0,\tau)\right|^2$
and therefore vanishes (see (\ref{theta11iszero}))
and $h(\tau,\bar \tau)$ is
singular because of (\ref{bosonicGdelta}). Suitable regularization
gives
\be
\left|\theta\left[\hspace*{-6pt}\begin{array}{c}1/2\\
1/2\end{array}\hspace*{-6pt}\right](0,\tau)\right|^2 h(\tau,\bar
\tau)= C\cdot\left|\left.\partial_v\theta\left[\hspace*{-6pt}
\begin{array}{c}1/2\\
1/2\end{array}\hspace*{-6pt}\right](v,\tau)\right|_{v=0}\right|^2
=C\cdot4\pi^2\left|\eta(\tau)\right|^6,
\ee
where $C$ is a real constant. With (\ref{formulaforB2}) we finally
arrive at
\be
\delta_{\mathcal{T}}=-\pi^2(\log 3) g_s^2 B_2CC_m.
\ee
One the other hand we have (\ref{Leffoneloop}), i.e.
\be
\delta_{\mathcal{T}}=\frac{B_2}{6\pi}.
\ee
This fixes the matching coefficient
\be
C_m=-\frac{1}{6\pi^3(\log 3) g_s^2C}.\label{fixinggs}
\ee

\section{Non-compact orientifolds with $\Omega J$ projection}
\label{sectionorientifold} 

In this section we consider D-branes in orientifold models because
they are (as far as we know it today) among the best possibilities to
get a setup in superstring theory
that comes close to the standard model. We will work in
the non-compact case because as the matter fields, gauge fields and
gravity are localized on the D-branes we will not need to
compactify. This will also have the advantage that we will have more
models at our disposal. The $\mathbb{Z}_N$ orbifold action e.g. will
no longer have to act crystallographically and the number of D-branes
will not be fixed. We will consider a
non-standard orientifold projection in order to have D3-branes. For
simplicity we assume $N$ to be odd so that we only have D3-branes and
we will assume that the D3-branes are coincident and on top of O$3_{+}$-planes.
For the discussed orientifolds of $\mathbb{Z}_N$ orbifolds we have
$\mathcal{N}=1$ supersymmetry in $d=4$. After we review the partition
functions for annulus, Moebius strip and Klein bottle and find the
tadpole conditions in section \ref{sectiono11} we derive their
contributions to the one-loop renormalization of the Planck mass in
section \ref{sectiono12}. 

\subsection{Tadpole conditions}\label{sectiono11}

Let $\Omega$ be the world sheet parity transformation and
$J$ act on the transverse complex bosons
$Z^i=X^{2i+2}+iX^{2i+3},i=1,2,3$ as
\be
J\,Z^i=-Z^i.
\ee
We consider the $\Omega J$ orientifold of the non-compact
$\mathbb{Z}_N$ orbifold of type IIB superstring theory and we assume $N$ to
be odd. Therefore we only have D3-branes (for $N$ even we would also
have D7-branes). This model has been presented in \cite{Kakushadze98}
(see \cite{Kakushadze01} for a review). The one-loop amplitudes are
the torus $\mathcal{T}$, the annulus $\mathcal{A}$, the Moebius strip
$\mathcal{M}$ and the Klein bottle $\mathcal{K}$. The torus
contribution to the one-loop renormalization of the Planck mass is one 
half of the corresponding orbifold result. 
The $\mathcal{A}$, $\mathcal{M}$ and $\mathcal{K}$ partition
functions for the standard $\Omega$ orientifolds (that has only D9
branes if $N$ is odd) can e.g. be found in \cite{Ibanez98} (see also
\cite{Polchinski96} and \cite{AngelatonjSagnotti02}) and we use the
same convention as in \cite{Ibanez98} that we suppress the winding and
momentum sums. The annulus, Moebius and Klein bottle amplitudes
for the non-compact $\Omega J$ orientifolds have been presented in
\cite{{IbanezRabadanUranga98}}. 
Let us define $q=e^{2\pi i\tau}$. For the annulus
$\tau=\frac{1}{2} i\tau_2$.
The annulus partition function is given by
\ba
Z_{\mathcal{A}}=\frac{1}{4N}\int\limits_0^\infty\!\!\frac{d\tau_2}{\tau_2^3} 
\sum_{k=0}^{N-1} \textrm{Tr}\left[(1+(-1)^F)\theta^k q^{L_0}\right],
\label{annulusamplitude}
\ea
and we find
\ba
Z_{\mathcal{A}}&=&\frac{1}{4N}\int\limits_0^\infty\!\!\frac{d\tau_2}{\tau_2^3} 
\sum_{k=0}^{N-1}\sum_{\alpha,\beta=0,1} (-1)^{\alpha+\beta+\alpha\beta}
\frac{\theta\left[\hspace*{-6pt}\begin{array}{c}\alpha/2\\\beta/2\end{array}\hspace*{-6pt}\right]}{\eta^3}\no\\&&\qquad \times\prod_{i=1}^3 
\left|2\sin(\pi
kv_i)\right|\frac{\theta\left[\hspace*{-6pt}\begin{array}{c}\alpha/2\\\beta/2+kv_i\end{array}\hspace*{-6pt}\right]}{\theta\left[\hspace*{-6pt}\begin{array}{c}1/2\\1/2+kv_i\end{array}\hspace*{-6pt}\right]}\left(\textrm{Tr 
} \gamma_{k,3}\right)^2\label{annulusamplitudeintermideate}\\
&=&\frac{(1-1)}{4N}\int\limits_0^\infty\!\!\frac{d\tau_2} 
{\tau_2^3} \sum_{k=0}^{N-1}
\frac{\theta\left[\hspace*{-6pt}\begin{array}{c}0\\1/2\end{array}\hspace*{-6pt}\right](0,\frac{1}{2}i\tau_2)}{\eta(\frac{1}{2}i\tau_2)^3}\no\\&&\qquad \times\prod_{i=1}^3 
\left|2\sin(\pi
kv_i)\right|\frac{\theta\left[\hspace*{-6pt}\begin{array}{c}0\\1/2+kv_i\end{array}\hspace*{-6pt}\right](0,\frac{1}{2}i\tau_2)}{\theta\left[\hspace*{-6pt}\begin{array}{c}1/2\\1/2+kv_i\end{array}\hspace*{-6pt}\right](0,\frac{1}{2}i\tau_2)}\left(\textrm{Tr 
} \gamma_{k,3}\right)^2.\no\\
\label{annulusamplitudefinalD3}
\ea
For the Moebius amplitude we have
$\tau=\frac{1}{2}+\frac{1}{2}i\tau_2$. The Moebius partition function
is given by 
\ba
Z_{\mathcal{M}}=\frac{1}{4N}\int\limits_0^\infty\!\!\frac{d\tau_2}{\tau_2^3} 
\sum_{k=0}^{N-1} \textrm{Tr}\left[(1+(-1)^F)\Omega J\theta^k
q^{L_0}\right].
\label{moebiusdefinition}
\ea
If we let everything depend on
\be
q_{new}=q^2_{old}=e^{4i\pi\tau}=e^{-2\pi\tau_2},\label{q-new}
\ee
then we find
\ba
Z_{\mathcal{M}}
&=&\frac{(1-1)}{4N}\int\limits_0^\infty\!\!
\frac{d\tau_2}{\tau_2^3} \sum_{k=0}^{N-1}
\frac{\theta\left[\hspace*{-6pt}\begin{array}{c}1/2\\0\end{array}\hspace*{-6pt}\right](0,i\tau_2)\theta\left[\hspace*{-6pt}\begin{array}{c}0\\1/2\end{array}\hspace*{-6pt}\right](0,i\tau_2)}{\eta(i\tau_2)^3\theta\left[\hspace*{-6pt}\begin{array}{c}0\\0\end{array}\hspace*{-6pt}\right](0,i\tau_2)}\no\\&\times&\prod_{i=1}^3 
s_i(-2\sin(\pi kv_i))\frac{\theta\left[\hspace*{-6pt}\begin{array}{c}1/2\\kv_i\end{array}\hspace*{-6pt}\right](0,i\tau_2)\theta\left[\hspace*{-6pt}\begin{array}{c}0\\1/2+kv_i\end{array}\hspace*{-6pt}\right](0,i\tau_2)}{\theta\left[\hspace*{-6pt}\begin{array}{c}1/2\\1/2+kv_i\end{array}\hspace*{-6pt}\right](0,i\tau_2)\theta\left[\hspace*{-6pt}\begin{array}{c}0\\kv_i\end{array}\hspace*{-6pt}\right](0,i\tau_2)}\textrm{Tr 
} \gamma_{\Omega_k,3}^{-1}\gamma_{\Omega_k,3}^{T},\no\\
\label{moebiusamplitudefinalD3}
\ea
where $s_i=\textrm{sign}(\sin(2\pi kv_i))$.
For the Klein bottle we have $\tau=2i\tau_2$. The Klein bottle
partition function is given by
\ba
Z_{\mathcal{K}}=\frac{1}{4N}\int\limits_0^\infty\!\!\frac{d\tau_2}{\tau_2^3} 
\sum_{n,k=0}^{N-1} \textrm{Tr}\left[(1+(-1)^F)\Omega J\theta^k
q^{L_0(\theta^n)}\bar q^{\bar L_0(\theta^n)}\right]. \label{Kleinamplitude}
\ea
$\Omega$ exchanges $\theta^n$ with $\theta^{N-n}$. As we have chosen
$N$ to be odd only $n=0$ does survive in the trace. We arrive at
\ba
Z_{\mathcal{K}}&=&\frac{1}{4N}\int\limits_0^\infty\!\!\frac{d\tau_2}{\tau_2^3} 
\sum_{k=0}^{N-1}\sum_{\alpha,\beta=0,1} (-1)^{\alpha+\beta+\alpha\beta}
\frac{\theta\left[\hspace*{-6pt}\begin{array}{c}\alpha/2\\\beta/2\end{array}\hspace*{-6pt}\right]}{\eta^3}\no\\&&\qquad \times\prod_{i=1}^3 
\frac{\left|2\sin(2\pi kv_i)\right|}{4(\sin(\pi (kv_i+\frac{1}{2})))^2} 
\frac{\theta\left[\hspace*{-6pt}\begin{array}{c}\alpha/2\\\beta/2+2kv_i\end{array}\hspace*{-6pt}\right]}{\theta\left[\hspace*{-6pt}\begin{array}{c}1/2\\1/2+2kv_i\end{array}\hspace*{-6pt}\right]}
\label{Kleinamplitudeintermideate}\\
&=&\frac{(1-1)}{4N}\int\limits_0^\infty\!\!\frac{d\tau_2} 
{\tau_2^3} \sum_{k=0}^{N-1}
\frac{\theta\left[\hspace*{-6pt}\begin{array}{c}0\\1/2\end{array}\hspace*{-6pt}\right](0,2i\tau_2)}{\eta(2i\tau_2)^3}\no\\&&\qquad \times\prod_{i=1}^3 
\frac{\left|2\sin(2\pi kv_i)\right|}{4(\sin(\pi (kv_i+\frac{1}{2})))^2} 
\frac{\theta\left[\hspace*{-6pt}\begin{array}{c}0\\1/2+2kv_i\end{array}\hspace*{-6pt}\right](0,2i\tau_2)}{\theta\left[\hspace*{-6pt}\begin{array}{c}1/2\\1/2+2kv_i\end{array}\hspace*{-6pt}\right](0,2i\tau_2)}.
\ea
We show in appendix \ref{appendixtadpoles} that this leads to the
tadpole conditions
\ba
0&=&\frac{1}{4}\prod_{i=1}^3 \left|2\sin(\pi kv_i)\right|\left(\textrm{Tr
} \gamma_{k,3}\right)^2 + 2\prod_{i=1}^3 
s_i(-2\sin(\pi kv_i)) \textrm{Tr
}(\gamma_{\Omega_k,3}^{-1}\gamma_{\Omega_k,3}^{T}) \no\\&&+4 
\prod_{i=1}^3 \frac{\left|2\sin(2\pi kv_i)\right|}{4(\sin(\pi
(kv_i+\frac{1}{2})))^2}.\label{tadpolecondition1D3} 
\ea
that are equivalent to
\be
0=\left(\textrm{Tr }\gamma_{2k,3}\mp 4\prod_{i=1}^3 \frac{1}{2\cos(\pi
kv_i)}\right)^2.\label{tadpolecondition3D3}
\ee
As we are considering the non-compact case we have no untwisted
tadpole cancellation condition and the number of D3-branes that we
call $n_3$ is arbitrary. But we still have to impose the twisted
tadpole cancellation conditions $(k=1,\dots,N-1)$.\enlargethispage{1cm}
For $\mathbb{Z}_N$ orientifolds we have 
$\gamma_{k,3}=\gamma_{1,3}^k$, $k=1,\dots,N-1$, and
$\gamma_{1,3}^N=\textbf{1}$. 
Remember that $\sum\limits_{k=0}^{N-1}e^{2i\pi k/N}=0$.

\subsection{Contribution of $\mathcal{A}$, $\mathcal{M}$ and
$\mathcal{K}$ to the renormalization of the Planck mass}
\label{sectiono12} 

We generalize the results of \cite{Antoniadis96} and
\cite{Antoniadis02} that considered compactifications on K3 (and
therefore of the $\mathbb{Z}_2$ orientifold) to general non-compact
$\mathbb{Z}_N$ orientifolds with $N$ odd.\\

For $\mathcal{K}, \mathcal{A}, \mathcal{M}$ there is only one spin
structure. The even spin structure two graviton amplitude is given by
(see e.g. \cite{Forger96})
\ba
A^{(2)}&=&\sum_{(\alpha,\beta)=0,1}^{even}
\int\limits_0^\infty\!\frac{d\tau_2}{\tau_2^2}
(-)^{\alpha+\beta+\alpha\beta}  Z(\tau,\bar\tau, (\alpha,\beta))
\no\\&&\qquad\times \int\!d^2 z_{1}\int\!d^2 z_{2} \langle
V^{(0,0)}(z_1,\bar z_1) V^{(0,0)}(z_2,\bar z_2) \rangle_{
(\alpha,\beta)}.
\ea
The piece in second order in momentum will give us the one-loop
renormalization of the Planck mass. Neglecting terms proportional to 
$p_1^2, p_2^2, p_1\cdot p_2,
p_{1\mu} \varepsilon^{\mu\nu}_1, p_{2\rho}\varepsilon^{\rho\sigma}_2,
\eta_{\mu\nu}\varepsilon^{\mu\nu}_1,
\eta_{\rho\sigma}\varepsilon^{\rho\sigma}_2$ we find
\ba
&&\left.A^{(2)}(\tau,\bar\tau, (\alpha,\beta)) \right|_{O(p^2)} =\left. \int\!d^2 z_1 \int\!d^2 z_2 \langle V^{(0,0)}(z_1,\bar z_1) V^{(0,0)}(z_2,\bar z_2) \rangle_{ (\alpha,\beta)}\right|_{O(p^2)}\no\\&&
=\sum_{\sigma=\mathcal{K}, \mathcal{A}, \mathcal{M}}g_s^2 \int\!d^2 z_{1}\int\!d^2 z_{2} \varepsilon^1_{\mu\nu} \varepsilon^2_{\rho\sigma}p_2^\nu p_1^\sigma\eta^{\mu\rho}\Bigl[ \langle\partial X \partial X\rangle \langle \tilde\psi \tilde\psi\rangle^2_{(\alpha,\beta)}\no\\&&\qquad  -  \langle\bar\partial X \partial X\rangle \langle \psi \tilde\psi\rangle^2_{(\alpha,\beta)} +  \langle\bar\partial X \bar\partial X\rangle \langle \psi \psi\rangle^2_{(\alpha,\beta)}  -  \langle\partial X \bar\partial X\rangle \langle \tilde\psi \psi\rangle^2_{(\alpha,\beta)} \Bigl].
\ea
From now on we again measure everything in string units.
For $\mathcal{K}, \mathcal{A}, \mathcal{M}$ we have to act with the following involutions on the covering tori
\be
I_{\mathcal{A}}=I_{\mathcal{M}}=1-\bar z, \qquad I_{\mathcal{K}}=1-\bar z +\frac{\tau}{2}.
\ee
If on the covering torus we have $\langle\psi(z)\psi(w)\rangle_{\mathcal{T},(\alpha,\beta)}=P_F((\alpha,\beta);z,w)$ then by the method of images (see appendix of \cite{Antoniadis96})
\ba
\langle\psi(z)\psi(w)\rangle_{\sigma,(\alpha,\beta)}&=&P_F((\alpha,\beta);z,w)\\
\langle\psi(z)\tilde\psi(\bar w)\rangle_{\sigma,(\alpha,\beta)}&=&P_F((\alpha,\beta);z,I_{\sigma}(w))\\
\langle\tilde\psi(\bar z)\tilde\psi(\bar w)\rangle_{\sigma,(\alpha,\beta)}&=&\bar P_F((\bar\alpha,\bar\beta);\bar z,\bar w).
\ea
On the other hand
\be
 \left(\langle\psi(z)\psi(0)\rangle_{\mathcal{T},(\alpha,\beta)}\right)^2=-\partial^2_z\log\theta\left[\hspace*{-6pt}\begin{array}{c}1/2\\1/2\end{array}\hspace*{-6pt}\right](z,\tau)+4\pi i\partial_\tau\log\theta\left[\hspace*{-6pt}\begin{array}{c}\alpha/2\\\beta/2\end{array}\hspace*{-6pt}\right](0,\tau)\label{G^2}
\ee
i.e. it can be written as a sum of a term that is independent of the spin structure (but dependent on the position $z$ on the world-sheet) and therefore vanishes when summed over the spin structure (as the partition function vanishes due to supersymmetry) and a term independent of the position $z$ on the world-sheet (but dependent on the spin structure) that can be taken outside the world-sheet integral.
The surviving piece will be the same in $ \langle \tilde\psi
\tilde\psi\rangle^2_{(\alpha,\beta)}, \langle \psi
\tilde\psi\rangle^2_{(\alpha,\beta)},   \langle \psi
\psi\rangle^2_{(\alpha,\beta)},  \langle \tilde\psi
\psi\rangle^2_{(\alpha,\beta)}$ so we replace it by $ \langle
\psi\psi\rangle^2_{(\alpha,\beta)}$ everywhere.\\

The remaining integral over the bosonic correlators gives (see again
\cite{Antoniadis96})
\be
\int\!d^2 z_{1}\int\!d^2 z_{2} \Bigl[ \langle\partial X \partial
X\rangle -  \langle\bar\partial X \partial X\rangle +
\langle\bar\partial X \bar\partial X\rangle -  \langle\partial X
\bar\partial X\rangle \Bigl]=\left\{\begin{array}{l} \pi \tau_2/8
\textrm{ for } \sigma=\mathcal{A},\mathcal{M}\\ \pi \tau_2/2 \textrm{
for } \sigma=\mathcal{K} \end{array}\right. .
\ee
Using (\ref{definingdelta}) and (\ref{thetaheatequation}) we find for
the annulus
\ba
&&\hspace*{-1.5cm} \delta_{\mathcal{A}}=-g_s^2C_m \sum_{(\alpha,\beta)=0,1}^{even}  \int\limits_0^\infty\!\frac{d\tau_2}{\tau_2^2} (-)^{\alpha+\beta+\alpha\beta}  Z_{\mathcal{A}}(\tau,\bar\tau, (\alpha,\beta))
\frac{\left.\partial_v^2 \theta\left[\hspace*{-6pt}\begin{array}{c}\alpha/2\\\beta/2\end{array}\hspace*{-6pt}\right](v,\frac{1}{2}i\tau_2)\right|_{v=0}}{\theta\left[\hspace*{-6pt}\begin{array}{c}\alpha/2\\\beta/2\end{array}\hspace*{-6pt}\right](0,\frac{1}{2}i\tau_2)}\frac{\pi \tau_2}{8},\no\\
\ea
for the Moebius strip
\ba
&&\hspace*{-1.5cm} \delta_{\mathcal{M}}=-g_s^2C_m \sum_{(\alpha,\beta)=0,1}^{even}  \int\limits_0^\infty\!\frac{d\tau_2}{\tau_2^2} (-)^{\alpha+\beta+\alpha\beta}  Z_{\mathcal{M}}(\tau,\bar\tau, (\alpha,\beta))
\frac{\left.\partial_v^2 \theta\left[\hspace*{-6pt}\begin{array}{c}\alpha/2\\\beta/2\end{array}\hspace*{-6pt}\right](v,\frac{1}{2}i\tau_2+\frac{1}{2})\right|_{v=0}}{\theta\left[\hspace*{-6pt}\begin{array}{c}\alpha/2\\\beta/2\end{array}\hspace*{-6pt}\right](0,\frac{1}{2}i\tau_2+\frac{1}{2})}\frac{\pi \tau_2}{8}\no\\
\ea
and for the Klein bottle
\ba
&&\hspace*{-1.5cm} \delta_{\mathcal{K}}=-g_s^2C_m \sum_{(\alpha,\beta)=0,1}^{even}  \int\limits_0^\infty\!\frac{d\tau_2}{\tau_2^2} (-)^{\alpha+\beta+\alpha\beta}  Z_{\mathcal{K}}(\tau,\bar\tau, (\alpha,\beta))
\frac{\left.\partial_v^2 \theta\left[\hspace*{-6pt}\begin{array}{c}\alpha/2\\\beta/2\end{array}\hspace*{-6pt}\right](v,2i\tau_2)\right|_{v=0}}{\theta\left[\hspace*{-6pt}\begin{array}{c}\alpha/2\\\beta/2\end{array}\hspace*{-6pt}\right](0,2i\tau_2)}\frac{\pi \tau_2}{2}.\no\\
\ea
Using (\ref{theta11iszero}) to (\ref{theta1''}), the Riemann identity
(\ref{Riemmannidentity}) and the partition functions 
(\ref{annulusamplitudeintermideate}), (\ref{moebiusamplitudefinalD3})
and (\ref{Kleinamplitudeintermideate}) we get
for the annulus
\ba
\delta_{\mathcal{A}}&=&- g_s^2C_m\frac{1}{2N}\partial_v^2
\int\limits_0^\infty\!\!\frac{d\tau_2} {\tau_2^3} \sum_{k=1}^{N-1}
\frac{\theta\left[\hspace*{-6pt}\begin{array}{c}1/2\\1/2\end{array}\hspace*{-6pt}\right](\frac{v}{2},\frac{1}{2}i\tau_2)}{\eta(\frac{1}{2}i\tau_2)^3}\no\\&&\qquad \times\left.\left[\prod_{i=1}^3 
\left|2\sin(\pi kv_i)\right|
\frac{\theta\left[\hspace*{-6pt}\begin{array}{c}1/2\\1/2+kv_i\end{array}\hspace*{-6pt}\right](\frac{v}{2},\frac{1}{2}i\tau_2)}{\theta\left[\hspace*{-6pt}\begin{array}{c}1/2\\1/2+kv_i\end{array}\hspace*{-6pt}\right](0,\frac{1}{2}i\tau_2)}\right]\left(\textrm{Tr 
} \gamma_{k,3}\right)^2\frac{\pi\tau_2}{8}\right|_{v=0}\no\\&&
\label{deltaAstarting}
\ea
for the Moebius strip
\ba
\delta_{\mathcal{M}}&=&- g_s^2C_m\frac{1}{2N}\partial_v^2
\int\limits_0^\infty\!\!\frac{d\tau_2}{\tau_2^3} \sum_{k=1}^{N-1}
\frac{\theta\left[\hspace*{-6pt}\begin{array}{c}1/2\\1/2\end{array}\hspace*{-6pt}\right](\frac{v}{2},i\tau_2)\theta\left[\hspace*{-6pt}\begin{array}{c}0\\0\end{array}\hspace*{-6pt}\right](\frac{v}{2},i\tau_2)}{\eta(i\tau_2)^3\theta\left[\hspace*{-6pt}\begin{array}{c}0\\0\end{array}\hspace*{-6pt}\right](0,i\tau_2)}\no\\
&\times& \left[\prod_{i=1}^3 
s_i(-2\sin(\pi kv_i))
\frac{\theta\left[\hspace*{-6pt}\begin{array}{c}1/2\\1/2+kv_i\end{array}\hspace*{-6pt}\right](\frac{v}{2},i\tau_2)\theta\left[\hspace*{-6pt}\begin{array}{c}0\\kv_i\end{array}\hspace*{-6pt}\right](\frac{v}{2},i\tau_2)}{\theta\left[\hspace*{-6pt}\begin{array}{c}1/2\\1/2+kv_i\end{array}\hspace*{-6pt}\right](0,i\tau_2)\theta\left[\hspace*{-6pt}\begin{array}{c}0\\kv_i\end{array}\hspace*{-6pt}\right](0,i\tau_2)}\right]\no\\&\times&\left. 
\textrm{Tr }\gamma_{\Omega_k,3}^{-1}\gamma_{\Omega_k,3}^{T}\frac{\pi\tau_2}{8}
\right|_{v=0}\no\\&& \label{deltaMstarting}
\ea
and for the Klein bottle
\ba
\delta_{\mathcal{K}}&=&-g_s^2C_m\frac{1}{2N}\partial_v^2
\int\limits_0^\infty\!\!\frac{d\tau_2}{\tau_2^3} \sum_{k=1}^{N-1}
\frac{\theta\left[\hspace*{-6pt}\begin{array}{c}1/2\\1/2\end{array}\hspace*{-6pt}\right](\frac{v}{2},2i\tau_2)}{\eta(2i\tau_2)^3}\no\\
&&\qquad \times\left.\left[ \prod_{i=1}^3 
\frac{\left|2\sin(2\pi kv_i)\right|}{4(\sin(\pi (kv_i+\frac{1}{2})))^2}
\frac{\theta\left[\hspace*{-6pt}\begin{array}{c}1/2\\1/2+2kv_i\end{array}\hspace*{-6pt}\right](\frac{v}{2},2i\tau_2)}{\theta\left[\hspace*{-6pt}\begin{array}{c}1/2\\1/2+2kv_i\end{array}\hspace*{-6pt}\right](0,2i\tau_2)}\right]\frac{\pi\tau_2}{2}\right|_{v=0}.\no\\&&\label{deltaKstarting}
\ea
We go to the transverse channel.
For the annulus we have the transformations $t=\frac{1}{2}\tau_2,l=\frac{1}{t}$
\ba
\delta_{\mathcal{A}}&=&- g_s^2C_m\frac{1}{2^2}\frac{1}{2N}\partial_v^2
\int\limits_0^\infty\!\!dl \sum_{k=1}^{N-1}
\frac{(-i)e^{-\frac{\pi lv^2}{4}} 
\theta\left[\hspace*{-6pt}\begin{array}{c}1/2\\1/2\end{array}\hspace*{-6pt}\right](\frac{ivl}{2},il)}{\eta(il)^3}
\no\\&\times&\left.\left[\prod_{i=1}^3 \left|2\sin(\pi kv_i)\right|
\frac{e^{-\frac{\pi lv^2}{4}}\theta\left[\hspace*{-6pt}\begin{array}{c}1/2+kv_i\\1/2\end{array}\hspace*{-6pt}\right](\frac{ivl}{2},il)}{\theta\left[\hspace*{-6pt}\begin{array}{c}1/2+kv_i\\1/2\end{array}\hspace*{-6pt}\right](0,il)}\right]\left(\textrm{Tr } \gamma_{k,3}\right)^2\frac{\pi}{4l}\right|_{v=0}.\no\\&&
\ea
For the Moebius strip we have the transformations
$t=\frac{1}{\tau_2},l=\frac{t}{2}$
\ba
\delta_{\mathcal{M}}&=&- g_s^2C_m2\frac{1}{2N}\partial_v^2
\int\limits_0^\infty\!\!dl \sum_{k=1}^{N-1}
\frac{(-i)e^{-\frac{\pi lv^2}{2}}\theta\left[\hspace*{-6pt}\begin{array}{c}1/2\\1/2\end{array}\hspace*{-6pt}\right](ivl,2il)e^{-\frac{\pi lv^2}{2}}\theta\left[\hspace*{-6pt}\begin{array}{c}0\\0\end{array}\hspace*{-6pt}\right]](ivl,2il)}{\eta(2il)^3\theta\left[\hspace*{-6pt}\begin{array}{c}0\\0\end{array}\hspace*{-6pt}\right](0,2il)}\no\\&\times&\left[\prod_{i=1}^3 
s_i(-2\sin(\pi kv_i)) \frac{e^{-\frac{\pi
lv^2}{2}}\theta\left[\hspace*{-6pt}\begin{array}{c}1/2+kv_i\\1/2\end{array}\hspace*{-6pt}\right](ivl,2il)e^{-\frac{\pi
lv^2}{2}}\theta\left[\hspace*{-6pt}\begin{array}{c}kv_i\\0\end{array}\hspace*{-6pt}\right](ivl,2il)}{\theta\left[\hspace*{-6pt}\begin{array}{c}1/2+kv_i\\1/2\end{array}\hspace*{-6pt}\right](0,2il)\theta\left[\hspace*{-6pt}\begin{array}{c}kv_i\\0\end{array}\hspace*{-6pt}\right](0,2il)}\right]
\no\\&\times&\left.\textrm{Tr 
} \gamma_{\Omega_k,3}^{-1}\gamma_{\Omega_k,3}^{T}\frac{\pi}{16l}\right|_{v=0}.
\ea
For the Klein bottle we have the transformations $t=2\tau_2,l=\frac{1}{t}$
\ba
\delta_{\mathcal{K}}&=&-g_s^2C_m2^2\frac{1}{2N}\partial_v^2
\int\limits_0^\infty\!\!dl \sum_{k=1}^{N-1}
\frac{(-i)e^{-\frac{\pi lv^2}{4}}
\theta\left[\hspace*{-6pt}\begin{array}{c}1/2\\1/2\end{array}\hspace*{-6pt}\right](\frac{ivl}{2},il)}{\eta(il)^3}\no\\&
\times&\left.\left[ \prod_{i=1}^3
\frac{\left|2\sin(2\pi kv_i)\right|}{4(\sin(\pi (kv_i+\frac{1}{2})))^2}
\frac{e^{-\frac{\pi lv^2}{4}}\theta\left[\hspace*{-6pt}\begin{array}{c}1/2+2kv_i\\1/2\end{array}\hspace*{-6pt}\right](\frac{ivl}{2},il)}{\theta\left[\hspace*{-6pt}\begin{array}{c}1/2+2kv_i\\1/2\end{array}\hspace*{-6pt}\right](0,il)}\right]\frac{\pi}{4l}\right|_{v=0}.\no\\&&
\ea
Note that it is important here to take the derivatives with respect to 
$v$ only after one goes to the transverse channel.
Using (\ref{thetaderivative}) to (\ref{theta1''}) we arrive at
\ba
\delta_{\mathcal{A}}&=&-2\pi ig_s^2C_m\frac{1}{2^2}\frac{1}{2N}
\int\limits_0^\infty\!\!dl \sum_{k=1}^{N-1}\left(\frac{il}{2}\right)^2
\left[\prod_{j=1}^3 \left|2\sin(\pi kv_j)\right|\right]
\no\\&&\times \sum_{i=1}^3\left(i\pi+2i\pi
kv_i+f\left(\frac{1}{2}+kv_i,\frac{1}{2},il\right)\right)\left(\textrm{Tr
} \gamma_{k,3}\right)^2\frac{\pi}{4l}\\
\delta_{\mathcal{M}}&=&-2\pi ig_s^2C_m2\frac{1}{2N}
\int\limits_0^\infty\!\!dl \sum_{k=1}^{N-1}\left(il\right)^2
\left[\prod_{j=1}^3 s_j(-2\sin(\pi kv_j))\right]
\no\\&&\times \Bigr[f(0,0,2il)+\sum_{i=1}^3\left(i\pi+2i\pi
kv_i+f\left(\frac{1}{2}+kv_i,\frac{1}{2},2il\right)\right)\no\\&&+
\sum_{i=1}^3\left(2i\pi kv_i+f\left(kv_i,0,2il\right)\right)\Bigr]
\textrm{Tr } \gamma_{\Omega_k,3}^{-1}\gamma_{\Omega_k,3}^{T}
\frac{\pi}{16l}\\
\delta_{\mathcal{K}}&=&-2\pi ig_s^2C_m2^2\frac{1}{2N}
\int\limits_0^\infty\!\!dl \sum_{k=1}^{N-1}\left(\frac{il}{2}\right)^2
\left[\prod_{j=1}^3 \frac{\left|2\sin(2\pi kv_j)\right|}{4(\sin(\pi (kv_j+\frac{1}{2})))^2}\right]
\no\\&&\times \sum_{i=1}^3\left(i\pi+4i\pi
kv_i+f\left(\frac{1}{2}+2kv_i,\frac{1}{2},il\right)\right)\frac{\pi}{4l}.
\ea
The tadpole cancellation condition (\ref{tadpolecondition1D3}) garantees 
that the contribution of terms in the sum proportional to a constant
(see the $i\pi$) in
$\delta=\delta_{\mathcal{A}}+\delta_{\mathcal{M}}+\delta_{\mathcal{K}}$ 
vanishes. On the other hand $\sum\limits_{i=1}^3 v_i=0$ because we
consider supersymmetric models. All remaining terms in $\delta$ are
proportional to some $f(a,b,il)$. These functions (for
$a\in(-1/2,1/2)$) fall off rapidely
as $l\rightarrow\infty$. Therefore $\delta$ is free of ultraviolet
divergences due to tadpole cancellation. We are left with
\ba
\delta&=&\frac{i\pi^2}{16N}g_s^2C_m
\int\limits_0^\infty\!\!dl\, l \sum_{k=1}^{N-1}\Bigl\{
\frac{1}{2^2}\left[\prod_{j=1}^3 \left|2\sin(\pi kv_j)\right|\right]
\left(\textrm{Tr } \gamma_{k,3}\right)^2\no\\&\times&
\sum_{i=1}^3 f\left(\frac{1}{2}+kv_i,\frac{1}{2},il\right)
+ 2\left[\prod_{j=1}^3 s_j(-2\sin(\pi kv_j))\right]
\Bigr[f(0,0,2il)\no\\&&+\sum_{i=1}^3 f\left(\frac{1}{2}+kv_i,
\frac{1}{2},2il\right)+
\sum_{i=1}^3 f\left(kv_i,0,2il\right)\Bigr]
\textrm{Tr } \gamma_{\Omega_k,3}^{-1}\gamma_{\Omega_k,3}^{T}\no\\
&+&2^2\left[\prod_{j=1}^3 \frac{\left|2\sin(2\pi kv_j)\right|}{4(\sin(\pi (kv_j+\frac{1}{2})))^2}\right]
\sum_{i=1}^3 f\left(\frac{1}{2}+2kv_i,\frac{1}{2},il\right)\Bigr\}.
\ea
Explicitely using the tadpole cancellation conditions
(\ref{tadpolecondition3D3}) we arrive at 
\ba
\delta&=& \frac{i\pi^2}{4N}g_s^2C_m
\int\limits_0^\infty\!\!dl\, l \sum_{k=1}^{N-1}
\left[\prod_{j=1}^3 \left|\tan(\pi kv_j)\right| \right]\Bigl\{
\sum_{i=1}^3 f\left(\frac{1}{2}+kv_i,\frac{1}{2},il\right)\no\\&
-& 2\Bigr[\sum_{i=1}^3 f\left(\frac{1}{2}+kv_i,
\frac{1}{2},2il\right)+\sum_{i=1}^3 f\left(kv_i,0,2il\right)\Bigr]\no\\&
+&\sum_{i=1}^3 f\left(\frac{1}{2}+2kv_i,\frac{1}{2},il\right) \Bigr\}.
\label{deltacontribution}
\ea
Though the integral is free of ultraviolet divergences we may still
have some infrared divergences that can only by handled by considering
the Wilsonian couplings. $C_m$ is given by (\ref{fixinggs}). For the
non-compact $\mathbb{Z}_N$ orientifolds with odd $N$ we can estimate from
(\ref{deltacontribution}) the large $N$ behavior
\be
\delta=\delta_{\mathcal{A}}+\delta_{\mathcal{M}}+\delta_{\mathcal{K}} 
\stackrel{N\rightarrow\infty}{\longrightarrow}O(1)
\ee
that is subleading as compared to the torus contribution. By using
$\sum\limits_{i=1}^3 v_i=0$ and the properties (\ref{fproperty1}) to 
(\ref{fproperty3}) one can actually check that the contribution from
the sector $(N-k)$ cancels the contribution from the sector $k$, i.e.
\be
\delta=\delta_{\mathcal{A}}+\delta_{\mathcal{M}}+\delta_{\mathcal{K}} 
=0.
\ee
The one-loop renormalization of the four-dimensional Planck mass comes 
only from the torus that is given by one half of the orbifold result
(\ref{largeNtoruscontribution}) and is $O(N)$.

\section{Conclusions} \label{sectionconclusions} 

We have considered D-branes in orientifold models because
they are (as far as we know it today) among the best possibilities to
get a setup in superstring theory that comes close to the standard
model. We focused on the non-compact case because in these models
the matter fields, gauge fields and gravity are localized on the
D-branes and we do not need to compactify. The issue was to show that
the one-loop correction of the Planck mass can be arbitrary large in
string units. It is therefore possible to accommodate the measured
four-dimensional Planck mass as a one-loop effect and to have a string
scale far below the Planck scale.\\

To be more precise we have constructed non-compact orientifolds of
$\mathbb{Z}_N$ orbifolds of type IIB with induced gravity on
coincident D3-branes that are on top of O$3_{+}$-planes. That we consider
orientifolds of $\mathbb{Z}_N$ orbifolds is because they have
localized twisted sectors and therefore localized gravity. As we
consider the non-compact case the orbifold need not act
crystallographically. That we assumed $N$ to be odd and the 
D3-branes to be coincident and on top of O3$_{+}$-planes was just for
simplicity.\\

We have shown for the $\mathbb{Z}_N$ orientifolds
with odd $N$ that the contribution to the one-loop renormalization of
the four-dimensional Planck mass comes only from the torus and is
$O(N)$ as the contributions from annulus, Moebius strip and Klein bottle
cancel. The idea that four-dimensional gravity may be induced by
quantum corrections at the one-loop level can therefore be realized by
considering sufficiently large $N$.\\

Obviously the models presented in this paper are only toy models of
orientifold realizations of the standard model and
there is plenty of room for generalization. The aim will be
to construct more realistic brane induced gravity models that come
closer to (supersymmetric generalizations) of the standard model by
considering e.g. more general D-brane configurations, Scherk-Schwarz
directions or Wilson lines. One will have to check the higher
loop corrections to the Planck mass and also the renormalization of
higher derivative terms as e.g. the $R^2$-terms.

\section{Acknowledgments}

The author thanks Elias Kiritsis for suggesting the problem, for
discussions and helpful remarks and Emilian Dudas and Pierre Vanhove
for discussions. This work was supported in part by CPHT through EEC
contracts HPRN-CT-2000-00122, HPRN-CT-2000-00131 and
HPRN-CT-2000-00148 and in part by the Austrian FWF project
No. J2259-N02.

\appendix
\section{Theta functions}

We use the definitions of \cite{Polchinskibook}
\ba
&&\theta\left[\hspace*{-6pt}\begin{array}{c}a\\
b\end{array}\hspace*{-6pt}\right](z,\tau)
=\sum_{n=-\infty}^\infty\exp\left[i\pi(n+a)^2\tau+2\pi
i(n+a)(z+b)\right]\\
&&\eta(\tau)=q^{1/24}\prod_{n=1}^\infty (1-q^n).
\ea
This gives
\ba
&&\theta\left[\hspace*{-6pt}\begin{array}{c}-a\\
-b\end{array}\hspace*{-6pt}\right](z,\tau)=
\theta\left[\hspace*{-6pt}\begin{array}{c}a\\
b\end{array}\hspace*{-6pt}\right](-z,\tau)\label{thetaproperty1}\\
&&\theta\left[\hspace*{-6pt}\begin{array}{c}a+1\\
b\end{array}\hspace*{-6pt}\right](z,\tau)=
\theta\left[\hspace*{-6pt}\begin{array}{c}a\label{thetaproperty2}\\
b\end{array}\hspace*{-6pt}\right](z,\tau)\\
&&\theta\left[\hspace*{-6pt}\begin{array}{c}a\label{thetaproperty3}\\
b+1\end{array}\hspace*{-6pt}\right](z,\tau)=
e^{2i\pi a}\theta\left[\hspace*{-6pt}\begin{array}{c}a\\
b\end{array}\hspace*{-6pt}\right](z,\tau).
\ea
We have the modular transformations
\ba
&&\theta\left[\hspace*{-6pt}\begin{array}{c}a\\
b\end{array}\hspace*{-6pt}\right](z,\tau+1)=e^{-i\pi(a^2+a)}
\theta\left[\hspace*{-6pt}\begin{array}{c}a\\
\frac{1}{2}+a+b\end{array}\hspace*{-6pt}\right](z,\tau)
\label{T-transformations}\\
&&\theta\left[\hspace*{-6pt}\begin{array}{c}a\\
b\end{array}\hspace*{-6pt}\right](\frac{z}{\tau},-\frac{1}{\tau})
=(-i\tau)^{1/2}\exp\left[2\pi iab+\frac{i\pi z^2}{\tau}\right]
\theta\left[\hspace*{-6pt}\begin{array}{c}b\\
-a\end{array}\hspace*{-6pt}\right](z,\tau)\label{S-transformation}\\
&& \eta(\tau+1)=e^{\frac{i\pi}{12}}\eta(\tau)\\
&& \eta(-\frac{1}{\tau})=(-i\tau)^{1/2}\eta(\tau),
\ea
where the second property is shown using Poisson resummation
\be
\sum_{n=-\infty}^\infty\exp\left[-\pi an^2+2\pi
ibn\right]=a^{-1/2}\sum_{m=-\infty}^\infty\exp\left[-\frac{\pi(m-b)^2}{a}
\right].\label{Poissonresummation}
\ee
The theta-functions have the product representation
\be
\theta\left[\hspace*{-6pt}\begin{array}{c}a\\
b\end{array}\hspace*{-6pt}\right](v,\tau)=
e^{2i\pi ab}q^{\frac{a^2}{2}}z^a
\prod_{m=1}^\infty(1-q^m)\left(1+e^{2\pi ib}z
q^{m-\frac{1}{2}+a}\right) \left(1+e^{-2\pi ib}z^{-1}
q^{m-\frac{1}{2}-a}\right),\label{thetaasproduct}
\ee
where $q=e^{2i\pi\tau},z=e^{2i\pi v}$.
In particular we have
\be
\frac{\theta\left[\hspace*{-6pt}\begin{array}{c}a\\
b\end{array}\hspace*{-6pt}\right](0,\tau)}{\eta(\tau)}=
e^{2\pi iab}q^{\frac{a^2}{2}-\frac{1}{24}}\prod_{m=1}^\infty \left[1+e^{-2\pi ib}q^{(m-\frac{1}{2}-a)}\right]\left[1+e^{2\pi ib}q^{(m-\frac{1}{2}+a)}\right].
\ee
For the derivatives we get
\be
\left.\frac{\partial_v\theta\left[\hspace*{-6pt}\begin{array}{c}a\\
b\end{array}\hspace*{-6pt}\right](v,\tau)}
{\theta\left[\hspace*{-6pt}\begin{array}{c}a\\
b\end{array}\hspace*{-6pt}\right](v,\tau)}\right|_{v=0}=2i\pi
a+f(a,b;\tau),\label{thetaderivative}
\ee
where
\be
f(a,b;\tau)=2\pi i\sum_{m=1}^\infty\left[\frac{e^{2i\pi
b}q^{m-\frac{1}{2}+a}}{1+e^{2i\pi
b}q^{m-\frac{1}{2}+a}}-\frac{e^{-2i\pi
b}q^{m-\frac{1}{2}-a}}{1+e^{-2i\pi b}q^{m-\frac{1}{2}-a}}\right].
\ee
From the behaviour of the theta functions follows
\ba
&&f(-a,-b;\tau)=-f(a,b;\tau)\label{fproperty1}\\
&&f(a+1,b;\tau)=-2i\pi+f(a,b;\tau)\label{fproperty2}\\
&&f(a,b+1;\tau)=f(a,b;\tau)\label{fproperty3}\\
&&f(a,b;\tau+1)=f(a,1/2+a+b;\tau)\label{fproperty4}\\
&&f(a,b;-1/\tau)=2i\pi(b-a)+f(b,-a;\tau)\label{fproperty5}.
\ea
The theta functions are solutions of the heat equation
\be
\frac{\partial^2}{\partial z^2}\theta\left[\hspace*{-6pt}\begin{array}{c}a\\b\end{array}\hspace*{-6pt}\right](z,\tau)=4\pi i\frac{\partial}{\partial\tau}\theta\left[\hspace*{-6pt}\begin{array}{c}a\\b\end{array}\hspace*{-6pt}\right](z,\tau)\label{thetaheatequation}
\ee
moreover
\ba
&&\theta\left[\hspace*{-6pt}\begin{array}{c}1/2\\1/2\end{array}\hspace*{-6pt}\right](0,\tau)=0\label{theta11iszero}\\
&&\left.\partial_v\theta\left[\hspace*{-6pt}\begin{array}{c}1/2\\1/2\end{array}\hspace*{-6pt}\right](v,\tau)\right|_{v=0}=-2\pi\eta(\tau)^3 \label{theta11and eta}\\
&&\left.\partial_v^2\theta\left[\hspace*{-6pt}\begin{array}{c}1/2\\1/2\end{array}\hspace*{-6pt}\right](v,\tau)\right|_{v=0}=0\label{theta1''}.
\ea
For $\sum\limits_{i=1}^4 h_i=\sum\limits_{i=1}^4 g_i=0$ we have the Riemann identity
\ba
&&\frac{1}{2}\sum_{\alpha,\beta=0}^{1}(-)^{\alpha+\beta+\alpha\beta} \prod_{i=1}^4 \theta\left[\hspace*{-6pt}\begin{array}{c}\alpha/2+h_i\\\beta/2+g_i\end{array}\hspace*{-6pt}\right] (v_i)=-\prod_{i=1}^4 \theta\left[\hspace*{-6pt}\begin{array}{c}1/2-h_i\\1/2-g_i\end{array}\hspace*{-6pt}\right] (v'_i) \label{Riemmannidentity} \\
&& v'_1=\frac{1}{2}(-v_1+v_2+v_3+v_4),\qquad v'_2=\frac{1}{2}(v_1-v_2+v_3+v_4)\\
&& v'_3=\frac{1}{2}(v_1+v_2-v_3+v_4),\qquad v'_4=\frac{1}{2}(v_1+v_2+v_3-v_4).
\ea

\section{The normalization of the partition function}
\label{Z3example}
 
\subsection{The example of the $\mathbb{Z}_3$ orbifold} 

Let us first consider the compact case.

\subsubsection{The massless spectrum}

Let $T^6=T^2\times T^2\times T^2$ and let us define
$\alpha=e^{\frac{2\pi i}{3}}$. The ${\mathbb Z}_3$ orbifold acts on
the tori as $Z^i\rightarrow e^{2\pi i v_i}Z^i$, where
$v=(v_1,v_2,v_3)=(\frac{1}{3},\frac{1}{3},-\frac{2}{3})$, i.e. we have
the reflection
\be
r: Z^1\rightarrow \alpha Z^1, Z^2\rightarrow \alpha Z^2,
Z^3\rightarrow \alpha^{-2} Z^3.
\ee
For the orbifold to act crystallographically the torus moduli of the
three spacetime tori have to be $\tau_i=\alpha^{1/2}R_i$. For each
torus we have 3 fixed points at $n\alpha^{1/4}R_i/3$, $n=0,1,2$,
giving a total of $3^3=27$ fixed points.\\

\textbf{The untwisted massless spectrum}\\

The zero point energy of a complex boson with twist $\theta$ is
\be
f(\theta)=\frac{1-3(1-2\theta)^2}{24}\label{zeropointenergy}
\ee
and the negative of this for a complex fermion. The shift is zero for
the 4 complex bosons (the transverse and the three compact) so the
zero point energy is in the NS sector
\be
4f(0)-4f(1/2)=-\frac{1}{2}
\ee
and the first exited states are massless. In the R sector instead we
have the zero point energy
\be
4f(0)-4f(0)=0
\ee
and the massless states are the degenerate ground states.\\

Let us seperate the lefthanded part of the massless states
acording to their eigenvalue under $\alpha$:
\ba
&&\alpha^0:
\psi^\mu_{-1/2}\left|0\right\rangle_{NS},\left|\frac{1}{2},\textbf{1}\right\rangle_R,\left|-\frac{1}{2},\bar\textbf{1}\right\rangle_R\\
&&\alpha^1:\psi^i_{-1/2}\left|0\right\rangle_{NS},\left|\frac{1}{2},\textbf{3}\right\rangle_R\\
&&\alpha^2:\psi^{\bar
i}_{-1/2}\left|0\right\rangle_{NS},\left|-\frac{1}{2},\bar\textbf{3}\right\rangle_R,
\ea
where
\ba
&&\left|\textbf{1}\right\rangle_R=
\left|\frac{1}{2},\frac{1}{2},\frac{1}{2}\right\rangle_R\\
&&\left|\bar\textbf{1}\right\rangle_R=
\left|-\frac{1}{2},-\frac{1}{2},-\frac{1}{2}\right\rangle_R\\
&&\left|\textbf{3}\right\rangle_R=
\Bigl\{\left|\frac{1}{2},-\frac{1}{2},-\frac{1}{2}\right\rangle_R,
\left|-\frac{1}{2},\frac{1}{2},-\frac{1}{2}\right\rangle_R,
\left|-\frac{1}{2},-\frac{1}{2},\frac{1}{2}\right\rangle_R\Bigr\}\\
&&\left|\bar\textbf{3}\right\rangle_R=
\Bigr\{\left|-\frac{1}{2},\frac{1}{2},\frac{1}{2}\right\rangle_R,
\left|\frac{1}{2},-\frac{1}{2},\frac{1}{2}\right\rangle_R,
\left|\frac{1}{2},\frac{1}{2},-\frac{1}{2}\right\rangle_R\Bigr\}.
\ea

The untwisted massless states that are invariant under the orbifold
action come from $\alpha^0\alpha^0$, $\alpha^1\alpha^2$ and
$\alpha^2\alpha^1$. We find 44 bosonic states
and 44 fermionic states
that give the following ${\mathcal N}=2$ multiplets
\ba
&&\Bigl[\Bigl(-2,-\frac{3}{2}^2,-1\Bigr)+\Bigl(1,\frac{3}{2}^2,2\Bigr)\Bigr]+\Bigl[\Bigl(-1,-\frac{1}{2}^2,0\Bigr)+\Bigl(0,\frac{1}{2}^2,1\Bigr)\Bigr]^9\no\\&+&\Bigr[\Bigl(-\frac{1}{2},0^2,\frac{1}{2}\Bigr)+\Bigl(-\frac{1}{2},0^2,\frac{1}{2}\Bigr)\Bigr],\label{untwistedspectrumZ3}
\ea
where the superscripts are not powers but give the number of fields of 
given helicity.\\

\textbf{The twisted massless spectrum}\\

The massless spectrum is 27 copies of the massless spectrum at one of
the fixed points. Let us first consider the states twisted by $r$.
The transverse complex bosons has shift 0 and the three compact
complex bosons have shift $1/3$. In the NS sector we get using
(\ref{zeropointenergy}) the zero point energy 
\be
f(0)-f(1/2)+3f(1/3)-3f(1/6)=0
\ee
and in the R sector
\be
f(0)-f(0)+3f(1/3)-3f(1/3)=0
\ee
so in both cases the massless states are ground states. Let us first
concider the R case. There are 2 fermion zero modes comming from the
transverse complex fermion leading to 2 possible states
$\left|\pm\frac{1}{2}\right\rangle_{h,R}$. The GSO
projection then only leaves only one state
$\left|\frac{1}{2}\right\rangle_{h,R}$. In the NS case there is a
unique ground state $\left|0\right\rangle_{h,NS}$. We find two bosonic 
states and two fermionic states.
The states twisted by $r^2$ give the antiparticles of these. The total
massless spectrum from the twisted states is in terms of ${\mathcal
N}=2$ multiplets
\be
\Bigl[\Bigl(-1,-\frac{1}{2}^2,0\Bigr)+\Bigl(0,\frac{1}{2}^2,1\Bigr)\Bigr]^{27}.
\label{twistedspectrumZ3}
\ee

\subsubsection{The helicity generating partition function}

For type IIB on $M^4\times T^6/{\mathbb Z}_3$ with $T^6=T^2\times
T^2\times T^2$ we find
\ba
&&\hspace*{-2cm}Z^{(0,0)}(v,\bar v)=N_0(N)\int\limits_{\mathcal
F}\!\frac{d^2\tau}{\tau_2^2}Z_X^2(\tau)
\left.\left[Z_\psi^+(v,\tau)Z_\psi^+(v,\tau)^*\right]\right|_{h=g=0}
\prod_{i=1}^3 Z_i\left[\hspace*{-6pt}\begin{array}{c}0\\0
\end{array}\hspace*{-6pt}\right](\tau) \label{Z3partitionfunction1}\\
&&\hspace*{-2cm}Z'(v,\bar v)=N_0(N)\int\limits_{\mathcal
F}\!\frac{d^2\tau}{\tau_2^2}Z_X^2(\tau)
\hspace*{-1cm}\sum_{\begin{array}{c}
h,g=0\\(h,g)\not=(0,0)\end{array}}^{N-1}\hspace*{-1cm}
Z_\psi^+(v,\tau)Z_\psi^+(v,\tau)^*
\prod_{i=1}^3 Z_i\left[\hspace*{-6pt}\begin{array}{c}hv_i\\gv_i
\end{array}\hspace*{-6pt}\right](\tau),\label{Z3partitionfunction}
\ea
\vspace{-1cm}\\where
\ba
Z_X^2(\tau)&=&\frac{1}{\tau_2}\frac{1}{|\eta(\tau)|^4}\\
Z_i\left[\hspace*{-6pt}\begin{array}{c}0\\0\end{array}\hspace*{-6pt}\right]
(\tau)&=&\frac{\Gamma_{2,2}}{|\eta(\tau)|^4}\ 
(\Gamma_{2,2}\textrm{ is the }(2,2)\textrm{ lattice sum})\\
Z_i\left[\hspace*{-6pt}\begin{array}{c}p\\q\end{array}\hspace*{-6pt}\right]
(\tau)&=&-3\left|\frac{\eta(\tau)}{\theta\left[\hspace*{-6pt}\begin{array}
{c}1/2+p\\1/2+q\end{array}\hspace*{-6pt}\right](0,\tau)}\right|^2\textrm{ 
for }(p,q)\not=(0,0)\label{Z3twistedboson}\\
Z_\psi^+(v,\tau)&=&\frac{\xi(v)}{2}\frac{1}{\eta(\tau)^4}
\sum_{\alpha,\beta=0}^{1} (-)^{\alpha+\beta+\alpha\beta}
\theta\left[\hspace*{-6pt}\begin{array}{c}\alpha/2\\
\beta/2\end{array}\hspace*{-6pt}\right](v,\tau)
\theta\left[\hspace*{-6pt}\begin{array}{c}\alpha/2+h/3\\\beta/2+g/3\end{array}
\hspace*{-6pt}\right](0,\tau)\no\\&&\times
\theta\left[\hspace*{-6pt}\begin{array}{c}\alpha/2+h/3\\\beta/2+g/3\end{array}
\hspace*{-6pt}\right](0,\tau)
\theta\left[\hspace*{-6pt}\begin{array}{c}\alpha/2-2h/3\\\beta/2-2g/3
\end{array}\hspace*{-6pt}\right](0,\tau).
\ea
Using the Riemann identity we get
\be\hspace*{-2cm}
Z_\psi^+(v,\tau)=\frac{\xi(v)}{\eta(\tau)^4}
\theta\left[\hspace*{-6pt}\begin{array}{c}1/2\\
1/2\end{array}\hspace*{-6pt}\right](v/2,\tau)
\left(\theta\left[\hspace*{-6pt}\begin{array}{c}1/2-h/3\\
1/2-g/3\end{array}\hspace*{-6pt}\right](v/2,\tau)\right)^2
\theta\left[\hspace*{-6pt}\begin{array}{c}1/2+2h/3\\
1/2+2g/3\end{array}\hspace*{-6pt}\right](v/2,\tau).
\ee
We find the contribution to the helicity generating partition
function from the massless modes from the limit
$\tau_2\rightarrow\infty$. The twisted states come from $h=1,2$ and the
untwisted from $h=0$. Using
\ba
&&\left.\Gamma_{2,2}\right|_{\textrm{massless}}=1,\qquad
\xi(v)\stackrel{\tau_2\rightarrow \infty}{\longrightarrow}1,\qquad
\bar\xi(\bar v)\stackrel{\tau_2\rightarrow
\infty}{\longrightarrow}1\label{masslesslimit1}\\ 
&&\frac{1}{|\eta(\tau)|^6}\theta\left[\hspace*{-6pt}\begin{array}{c}1/2\\
1/2\end{array}\hspace*{-6pt}\right](v/2,\tau)\bar\theta\left[\hspace*{-6pt}
\begin{array}{c}1/2\\1/2\end{array}\hspace*{-6pt}\right](\bar v/2,\tau)=
4\sin\frac{\pi v}{2}\sin\frac{\pi \bar v}{2} (1+O(q\bar
q))\no\\&&\label{masslesslimit2}\\ 
&&\int\limits_{\mathcal F}\!\frac{d^2\tau}{\tau_2^3}=\log
3\label{masslesslimit3} 
\ea
and
\ba
&&\left|\frac{\theta\left[\hspace*{-6pt}\begin{array}{c}1/2\\1/6\end{array}
\hspace*{-6pt}\right](v/2,\tau)}{\theta\left[\hspace*{-6pt}\begin{array}{c}
1/2\\1/6\end{array}\hspace*{-6pt}\right](0,\tau)}\right|\stackrel{\tau_2
\rightarrow\infty}{\longrightarrow}\frac{1}{\sqrt{3}}\left|e^{i\pi
v/2}+e^{-i\pi/3}e^{-i\pi v/2}\right|\label{Z3thetalimit1}\\
&&\left|\frac{\theta\left[\hspace*{-6pt}\begin{array}{c}1/2\\5/6\end{array}
\hspace*{-6pt}\right](v/2,\tau)}{\theta\left[\hspace*{-6pt}\begin{array}{c}
1/2\\5/6\end{array}\hspace*{-6pt}\right](0,\tau)}\right|\stackrel{\tau_2
\rightarrow\infty}{\longrightarrow}\frac{1}{\sqrt{3}}\left|e^{-i\pi/3}e^{i\pi
v/2}+e^{-i\pi v/2}\right|\\
&&\left|\frac{\theta\left[\hspace*{-6pt}\begin{array}{c}1/6\\b\end{array}
\hspace*{-6pt}\right](v/2,\tau)}{\theta\left[\hspace*{-6pt}\begin{array}{c}
1/6\\b\end{array}\hspace*{-6pt}\right](0,\tau)}\right|\stackrel{\tau_2
\rightarrow\infty}{\longrightarrow}\left|e^{i\pi v/6}\right|\\
&&\left|\frac{\theta\left[\hspace*{-6pt}\begin{array}{c}5/6\\b\end{array}
\hspace*{-6pt}\right](v/2,\tau)}{\theta\left[\hspace*{-6pt}\begin{array}{c}
5/6\\b\end{array}\hspace*{-6pt}\right](0,\tau)}\right|\stackrel{\tau_2
\rightarrow\infty}{\longrightarrow}\left|e^{-i\pi v/6}\right|
\label{Z3thetalimit2}
\ea
that follows from the product representation of the theta functions
(\ref{thetaasproduct})
we find the massless contribution of the twisted sector
\ba
\left.Z^T(v,\bar v)\right|_{\textrm{massless}}&=&3 N_0\ (-3)^34
\left|\sin\frac{\pi v}{2}\right|^2\left|e^{i\pi v/6}\right|^6\no\\&+&3
N_0\ (-3)^34\left|\sin\frac{\pi v}{2}\right|^2\left|e^{-i\pi
v/6}\right|^6 \label{Z3partitionfunctionmasslesstwisted}
\ea
and the massless contribution of the untwisted sector
\ba
\left.Z^U(v,\bar v)\right|_{\textrm{massless}}&=& N_0\ 256
\left|\sin\frac{\pi v}{2}\right|^8\no\\&-& N_0\ 4
\left|\sin\frac{\pi v}{2}\right|^2\left|e^{i\pi
v/2}+e^{-i\pi/3}e^{-i\pi v/2}\right|^6\no\\&-& N_0\ 4
\left|\sin\frac{\pi v}{2}\right|^2\left|e^{-i\pi/3}e^{i\pi
v/2}+e^{-i\pi v/2}\right|^6.\no\\ \label{Z3partitionfunctionmasslessuntwisted}
\ea
If we write a function $f(v, \bar v)$ as
\be
f(v, \bar v)=\left(\sum_{\lambda_R}\tilde c_{\lambda_R}e^{2i\pi
v\lambda_R}\right) \left(\sum_{\lambda_L}c_{\lambda_L}e^{-2i\pi \bar
v\lambda_L}\right)
\ee
with coefficients $\tilde c_{\lambda_R}, c_{\lambda_L}$ then the contribution 
to the fixed helicity $\lambda_{\textrm{tot}}=\lambda_R+\lambda_L$ is
\be
\sum_{\lambda_R}\tilde c_{\lambda_R}c_{(\lambda_{tot}-\lambda_R)}
e^{2i\pi v\lambda_R} \left.e^{-2i\pi \bar v
(\lambda_{tot}-\lambda_R)}\right|_{v=\bar
v=0}=\sum_{\lambda_R}\tilde
c_{\lambda_R}c_{(\lambda_{tot}-\lambda_R)}.
\label{helicitycontentfixing}
\ee
The helicity content of $4\left|\sin\frac{\pi
v}{2}\right|^2\left|e^{i\pi v/6}\right|^6$ is
\begin{tabular}{r|ccc}
$\lambda_{\textrm{tot}}$ & 0 & 1/2 & 1\\
\hline\\
value & $-1$ & 2 & $-1$
\end{tabular}.\\
The helicity content of $4\left|\sin\frac{\pi
v}{2}\right|^2\left|e^{-i\pi v/6}\right|^6$ is
\begin{tabular}{r|ccc}
$\lambda_{\textrm{tot}}$ & 0 & $-1/2$ & $-1$\\
\hline\\
value & $-1$ & 2 & $-1$
\end{tabular}.\\
Comparing with the twisted spectrum (\ref{twistedspectrumZ3}) fixes the
normalization to be
\be
N_0=\frac{1}{3}.\label{normalizationZ3}
\ee
The helicity content of $4\left|\sin\frac{\pi v}{2}\right|^2\left|e^{i\pi
v/2}+e^{-i\pi/3}e^{-i\pi v/2}\right|^6$ and of\\ $4\left|\sin\frac{\pi
v}{2}\right|^2\left|e^{-i\pi/3}e^{i\pi v/2}+e^{-i\pi v/2}\right|^6$
both give\\
\hspace*{2.5cm}\begin{tabular}{r|ccccc}
$\lambda_{\textrm{tot}}$ & 0 & $\pm 1/2$ & $\pm 1$ &  $\pm 3/2$ & $\pm 2$\\
\hline\\
value & 2 & 2 & $-1$ & $-1$ & $-1$
\end{tabular}.\\
The helicity content of $256\left|\sin\frac{\pi v}{2}\right|^8$ is\\
\hspace*{2.5cm}\begin{tabular}{r|ccccc}
$\lambda_{\textrm{tot}}$ & 0 & $\pm 1/2$ & $\pm 1$ &  $\pm 3/2$ & $\pm 2$\\
\hline\\
value & 70 & $-56$ & 28 & $-8$ & 1
\end{tabular}.\\
We see that with the normalization (\ref{normalizationZ3}) we indeed
reproduce the untwisted spectrum (\ref{untwistedspectrumZ3}).\\

Let us also compute the second helicity supertrace
\be
B_2=-\left.\Bigl(\frac{1}{2\pi i}\partial_v-\frac{1}{2\pi i}\partial_{\bar
v}\Bigr)^2 Z(v,\bar v)\right|_{v=\bar v=0}.
\ee
Using
\be
\left.\partial_{\nu}\theta\left[\hspace*{-6pt}\begin{array}{c}1/2\\
1/2\end{array}\hspace*{-6pt}\right](\nu,\tau)\right|_{\nu=0}=-2\pi
\eta(\tau)^3,\qquad \frac{\partial}{\partial\nu}=\frac{1}{2}\frac{\partial}
{\partial\frac{\nu}{2}},
\ee
we find from (\ref{Z3partitionfunction1}) and (\ref{Z3partitionfunction})
\be
B_2=36.
\ee
The massless contribution from
(\ref{Z3partitionfunctionmasslesstwisted}) and
(\ref{Z3partitionfunctionmasslessuntwisted}) gives the same
\be
\left.B_2\right|_{\textrm{massless}}=27+9=36.
\ee
The ${\mathbb Z}_3$ orbifold is the singular limit of the
Eguchi-Hanson space $EH_3$ that is a Calabi-Yau 3-fold with $h^{1,1}=36$ and
$h^{2,1}=0$. We get form (\ref{Leffoneloop})
\be
\Delta{\mathcal
L}_{\textrm{eff}}^{\textrm{1-loop}}=\frac{6}{\pi} M_s^2 \sqrt{-g}R.
\ee

\subsubsection{The non-compact case}

For the ${\mathbb Z_3}$ orbifold we have in the non-compact case from the 27
fixed points of the compact case just the origin left. Instead of $C=-3$
in (\ref{Z3twistedboson}) we have $C=-1$. For the second helicity
supertrace we get
\be
B_2^T=\left.B_2^T\right|_{\textrm{massless}}=1
\ee
that gives using (\ref{Leffoneloop})
\be
\Delta{\mathcal L}_{\textrm{eff}}^{\textrm{1-loop}}
=\frac{1}{6\pi} M_s^2 \sqrt{-g}R.
\ee

\subsection{General $N$}

We consider the non-compact case and fix the normalization from the
twisted sectors. The massless spectrum is given by
(\ref{generaltwistedspectrum1}) or (\ref{generaltwistedspectrum2}),
where the sectors $h$ and $N-h$ (for $h\not=\frac{N}{2}$) together
give one vector multiplet as does the sector $h=\frac{N}{2}$ if $N$ is 
even.\\

We will proof that the normalization of the partition function is
given by (\ref{partitionfunctionnormalization}) in the case that $N$
is prime. The proof in the general case $N\in\mathbb{N}$ relies on the
fact that every natural number can uniquely be written as a product of
prime numbers and is quite lengthy as there are
sectors with $hv_i\in \mathbb{Z}$ and one has more cases to
consider. However, this generalization is straight forward. We will also
assume that $v_i\notin\mathbb{Z}$, $i=1,2,3$. The $v_i$ are of the
form $v_i=\frac{k_i}{N}$ with $k_i\in\mathbb{Z}$. As $h=1,\dots,N-1$
and $N$ is prime it follows that $hv_i\notin\mathbb{Z}$ for all $h$.\\

From the partition function (\ref{ZNZMpartitionfunction}) we get the
massless contribution of the twisted sectors from the limit
$\tau_2\rightarrow\infty$. Using (\ref{masslesslimit1}) to
(\ref{masslesslimit3}) and $C^{(N)}=-1$ we arrive at
\be
\left.Z^T(v,\bar v)\right|_{\textrm{massless}}=-N_0\ 4
\left|\sin\frac{\pi v}{2}\right|^2\sum_{g=0}^{N-1} \sum_{h=1}^{N-1} 
\prod_{i=1}^3 \lim_{\tau_2\rightarrow\infty} 
\left|\frac{\theta\left[\hspace*{-6pt}\begin{array}{c}1/2+hv_i\\
1/2+gv_i\end{array} \hspace*{-6pt}\right](v/2,\tau)}
{\theta\left[\hspace*{-6pt}\begin{array}{c}1/2+hv_i\\
1/2+gv_i\end{array} \hspace*{-6pt}\right](0,\tau)}\right|^2.
\ee
We write $hv_i=[hv_i]+r(hv_i)$ with integer part $[hv_i]$ and rest
$r(hv_i)\in(0,1)$. From the product representation of the theta
functions (\ref{thetaasproduct}) we get
\ba
&&\prod_{i=1}^3 \sum_{g=0}^{N-1} \sum_{h=1}^{N-1} 
\lim_{\tau_2\rightarrow\infty} 
\left|\frac{\theta\left[\hspace*{-6pt}\begin{array}{c}1/2+hv_i\\
1/2+gv_i\end{array} \hspace*{-6pt}\right](v/2,\tau)}
{\theta\left[\hspace*{-6pt}\begin{array}{c}1/2+hv_i\\
1/2+gv_i\end{array} \hspace*{-6pt}\right](0,\tau)}\right|^2=
\prod_{i=1}^3 \sum_{g=0}^{N-1} \sum_{h=1}^{N-1}
\left|e^{i\pi v\left(-\frac{1}{2}+r(hv_i)\right)}
\right|^2\no\\&=& N \sum_{h=1}^{N-1} \left|e^{-\frac{3}{2}i\pi v}
e^{i\pi vh\sum\limits_{i=1}^3 v_i} 
e^{-i\pi v\sum\limits_{i=1}^3 [hv_i]}\right|^2.
\ea
Due to supersymmetry we have $v_3=-v_1-v_2$. On the other hand
$[-x]=-[x]-1$ for any $x$ and $[x_1+x_2 ]=\left\{\begin{array}{c}
[x_1]+[x_2] \textrm{ for }r(x_1)+r(x_2)<1 \\
\textrm{[}x_1\textrm{]}+\textrm{[}x_2\textrm{]}+1 \textrm{ for
}r(x_1)+r(x_2)>1\end{array}\right.$ for any 
$x_1$ and $x_2$. If $r(hv_1)+r(hv_2)<1$ $(>1)$ then 
$r((N-h)v_1)+r((N-h)v_2)=1-r(hv_1)+1-r(hv_2)>1$ $(<1)$. We are left with
\be
\left.Z^T(v,\bar v)\right|_{\textrm{massless}}=-N_0\ 4
\left|\sin\frac{\pi v}{2}\right|^2 N \frac{N-1}{2} 
\left(\left|e^{-\frac{i\pi v}{2}}\right|^2+ 
\left|e^{\frac{i\pi v}{2}}\right|^2\right)
\ee
and from the helicity content of the functions after equation
(\ref{helicitycontentfixing}) follows the normalization
(\ref{partitionfunctionnormalization}) if one matches on the spectrum
(\ref{generaltwistedspectrum2}).

\section{Some details of the two graviton amplitude}
\label{appendixamplitudedetails}

The part of the general n-point one-loop amplitude comming from even-even spin structures is given by (see \cite{Forger96})
\ba
\mathcal{A}^{(e,e)}_n&=& \sum_{(\alpha,\beta)=0,1}^{even} \sum_{(\bar\alpha,\bar\beta)=0,1}^{even} \int\limits_{\tau\in\Gamma}\!\frac{d^2\tau}{\tau_2^2} (-)^{\alpha+\beta+\alpha\beta}(-)^{\bar\alpha+\bar\beta+\bar\alpha\bar\beta}  Z(\tau,\bar\tau, (\alpha,\beta), (\bar\alpha,\bar\beta)) \no\\&&\times \int\limits_{\Gamma_\tau}\prod_{i=1}^{n-1}d^2 z_i\langle\prod_{i=1}^n V^{(0,0)}(z_i,\bar z_i)\rangle_{ (\alpha,\beta), (\bar\alpha,\bar\beta)} \label{generalamplitude}
\ea
as all vertex operators can be chosen in the $(0,0)$-ghost picture (see e.g. \cite{Lerche87}), $\Gamma$ is the  fundamental region that is
\be
\Gamma=\{\tau| \textrm{Im}\tau>0,|\textrm{Re}\tau|\leq\frac{1}{2},|\tau|\geq 1\}
\ee
for the torus and $\tau_2\in [0,\infty] $ for $\mathcal{K}, \mathcal{A}, \mathcal{M}$ and the $z_i$ are integrated over the strip
\be
\Gamma_\tau=\{z_i|\,|\textrm{Re}z_i|\leq\frac{1}{2},0\leq \textrm{Im}z_i\leq\textrm{Im}\tau\}.
\ee
For the torus we can set $z_n=\tau$ due to the conformal symmetry.
The partition function is vanishing by supersymmetry so we need at least two fermion contractions (from the vertex operators) to get a non-vanishing result.
The graviton vertex operator in the $(0,0)$-ghost picture is
\be
V^{(0,0)}(z,\bar z)=-\frac{2g_s}{\alpha'}\, \varepsilon_{\mu\nu}\, :\,
\left(i\partial X^\mu-\frac{\alpha'}{2}\psi^\mu p\cdot\psi\right)
\left(i\bar\partial X^\nu+\frac{\alpha'}{2}\tilde\psi^\nu
p\cdot\tilde\psi\right) e^{ip\cdot X}:\, .
\ee
The bosonic Green function on the torus is
\be
\langle X^\mu(z,\bar z)X^\nu(z',\bar z')\rangle=-\frac{\alpha'}{2}\eta^{\mu\nu}\log|\chi(z-z',\tau)|^2,
\ee
where
\be
\chi(z_{ij},\tau)=2\pi \exp\left[-\pi\frac{(\textrm{Im}z_{ij})^2}{\textrm{Im}\tau}\right]\frac{\theta\left[\hspace*{-6pt}\begin{array}{c}1/2\\1/2\end{array}\hspace*{-6pt}\right](z_{ij},\tau)}{\theta'\left[\hspace*{-6pt}\begin{array}{c}1/2\\1/2\end{array}\hspace*{-6pt}\right](0,\tau)}.
\ee
The fermionic Green function on the torus is
\be
\langle\psi^\mu(z)\psi^\nu(z')\rangle_{(\alpha,\beta)} = \eta^{\mu\nu} \frac{\theta\left[\hspace*{-6pt}\begin{array}{c}\alpha/2\\\beta/2\end{array}\hspace*{-6pt}\right](z-z',\tau)\theta'\left[\hspace*{-6pt}\begin{array}{c}1/2\\1/2\end{array}\hspace*{-6pt}\right](0,\tau)}{\theta\left[\hspace*{-6pt}\begin{array}{c}1/2\\1/2\end{array}\hspace*{-6pt}\right](z-z',\tau)\theta\left[\hspace*{-6pt}\begin{array}{c}\alpha/2\\\beta/2\end{array}\hspace*{-6pt}\right](0,\tau)}.
\ee
For the bosonic correlation functions we use as a starting point
\ba
&&\langle\exp\Bigl[\sum_{i=1}^N \Bigl(ik_i\cdot X(z_i,\bar z_i)+J^\mu\partial_{z_i}X_\mu(z_i,\bar z_i)+\bar J^\mu\partial_{\bar z_i}X_\mu(z_i,\bar z_i)\Bigl)\Bigr]\rangle=\no\\&&=\exp\Bigl[\frac{1}{2}\sum_{i\not=j}^N\Bigl(ik_{i\mu}+J_{\mu}(z_i)\partial_{z_i}+\bar J_{\mu}(\bar z_i)\partial_{\bar z_i}\Bigr)\Bigl(ik_{j\nu}+J_{\nu}(z_j)\partial_{z_j}+\bar J_{\nu}(\bar z_j)\partial_{\bar z_j}\Bigr)\no\\&&\times\langle X^\mu(z_i,\bar z_i)X^\nu(z_j,\bar z_j)\rangle\Bigr].
\ea
Making functional derivatives with respect to the currents $J(z)$ and
$\bar J(\bar z)$ and finally setting them to zero, we can compute the
expectation value of any vertex operator that is a polynomial in
derivatives of $X$ times the exponential $e^{ik\cdot X}$.
We define
\be
G(z,\tau)=-\frac{1}{2}\log|\chi(z,\tau)|
\ee
and find
\begin{eqnarray}
\partial_z G(z,\tau)&=&-\frac{1}{4}\left.\sum_{k,m}\right.'\frac{1}{k\tau-m}\exp\Bigl[2\pi ik\Bigl(\textrm{Re}z-\textrm{Re}\tau\frac{\textrm{Im}z}{\textrm{Im}\tau}\Bigr)\Bigr]\exp\Bigl[2\pi im\frac{\textrm{Im}z}{\textrm{Im}\tau}\Bigr]\no\\
\partial_{\bar z} G(z,\tau)&=&\frac{1}{4}\left.\sum_{k,m}\right.'\frac{1}{k\bar\tau-m}\exp\Bigl[2\pi ik\Bigl(\textrm{Re}z-\textrm{Re}\tau\frac{\textrm{Im}z}{\textrm{Im}\tau}\Bigr)\Bigr]\exp\Bigl[2\pi im\frac{\textrm{Im}z}{\textrm{Im}\tau}\Bigr]\no\\
\partial_z\partial_z G(z,\tau)&=&\frac{\pi}{4\textrm{Im}\tau}\left.\sum_{k,m}\right.'\frac{m-k\bar\tau}{m-k\tau}\exp\Bigl[2\pi ik\Bigl(\textrm{Re}z-\textrm{Re}\tau\frac{\textrm{Im}z}{\textrm{Im}\tau}\Bigr)\Bigr]\exp\Bigl[2\pi im\frac{\textrm{Im}z}{\textrm{Im}\tau}\Bigr]\no\\
\partial_{\bar z}\partial_{\bar z} G(z,\tau)&=&\frac{\pi}{4\textrm{Im}\tau}\left.\sum_{k,m}\right.'\frac{m-k\tau}{m-k\bar\tau}\exp\Bigl[2\pi ik\Bigl(\textrm{Re}z-\textrm{Re}\tau\frac{\textrm{Im}z}{\textrm{Im}\tau}\Bigr)\Bigr]\exp\Bigl[2\pi im\frac{\textrm{Im}z}{\textrm{Im}\tau}\Bigr]\no\\
\partial_z\partial_{\bar z} G(z,\tau)&=&-\frac{\pi}{4\textrm{Im}\tau}\left.\sum_{k,m}\right.'\exp\Bigl[2\pi ik\Bigl(\textrm{Re}z-\textrm{Re}\tau\frac{\textrm{Im}z}{\textrm{Im}\tau}\Bigr)\Bigr]\exp\Bigl[2\pi im\frac{\textrm{Im}z}{\textrm{Im}\tau}\Bigr]\no\\&&=-\frac{\pi}{4}\Bigl(\delta(\textrm{Re}z)\delta(\textrm{Im}z)-\frac{1}{\textrm{Im}\tau}\Bigr),\label{bosonicGdelta}
\end{eqnarray}
where $\left.\sum_{k,m}\right.'$ means that $(k,m)=(0,0)$ is not in the sum. This gives
\ba
\int\limits_0^{\textrm{Im}\tau}\!\!d\textrm{Im}z\int\limits_{-\frac{1}{2}}^{\frac{1}{2}}\!\!d\textrm{Re}z\,\partial_z\partial_z G(z,\tau)&=&0\label{bosonicG2}\\
\int\limits_0^{\textrm{Im}\tau}\!\!d\textrm{Im}z\int\limits_{-\frac{1}{2}}^{\frac{1}{2}}\!\!d\textrm{Re}z\,\partial_z\partial_{\bar z} G(z,\tau)&=&0.\label{bosonicG3}
\ea

\section{Deriving the tadpole conditions} \label{appendixtadpoles}

First we find the transverse channel expressions for the
amplitudes. For the annulus and Klein bottle this is achieved by the
standard $S$-transformation as they depend on $\frac{1}{2}i\tau_2$ and 
$2i\tau_2$ respectively. For the Moebius amplitude the functions do
not depend on the standard $\frac{1}{2}+\frac{1}{2}i\tau_2$ that would 
leed to a $P$-transformation (with $P=ST^2ST$) but on $i\tau_2$
(because they depend on $q_{new}=q_{old}^2$ see (\ref{q-new})) so we
again need a $S$-transformation to go to the transverse channel. The
$S$-transformation of the theta functions is given by
(\ref{S-transformation}).\\

For the annulus we have the transformations $t=\frac{1}{2}\tau_2,l=\frac{1}{t}$
\ba
Z_{\mathcal{A}}
&=&\frac{1}{2^2}\frac{(1-1)}{4N}\int\limits_0^\infty\!\!dl \sum_{k=0}^{N-1}
\frac{\theta\left[\hspace*{-6pt}\begin{array}{c}1/2\\0\end{array}\hspace*{-6pt}\right](0,il)}{\eta(il)^3}\no\\&&\qquad \times\prod_{i=1}^3 
\left|2\sin(\pi kv_i)\right|
\frac{\theta\left[\hspace*{-6pt}\begin{array}{c}1/2+kv_i\\0\end{array}\hspace*{-6pt}\right](0,il)}{(-i)e^{-i\pi 
kv_i} \theta\left[\hspace*{-6pt}\begin{array}{c}1/2+kv_i\\1/2\end{array}\hspace*{-6pt}\right](0,il)}\left(\textrm{Tr } \gamma_{k,3}\right)^2.\no\\&&
\ea
For the Moebius strip we have the transformations $t=\frac{1}{\tau_2},l=\frac{t}{2}$
\ba
Z_{\mathcal{M}}
&=&2\frac{(1-1)}{4N}\int\limits_0^\infty\!\!
dl \sum_{k=0}^{N-1}
\frac{\theta\left[\hspace*{-6pt}\begin{array}{c}0\\1/2\end{array}\hspace*{-6pt}\right](0,2il)\theta\left[\hspace*{-6pt}\begin{array}{c}1/2\\0\end{array}\hspace*{-6pt}\right](0,2il)}{\eta(2il)^3\theta\left[\hspace*{-6pt}\begin{array}{c}0\\0\end{array}\hspace*{-6pt}\right](0,2il)}\no\\& \times&\prod_{i=1}^3 
s_i(-2\sin(\pi kv_i))\frac{e^{-i\pi
kv_i}\theta\left[\hspace*{-6pt}\begin{array}{c}kv_i\\1/2\end{array}\hspace*{-6pt}\right](0,2il)\theta\left[\hspace*{-6pt}\begin{array}{c}1/2+kv_i\\0\end{array}\hspace*{-6pt}\right](0,2il)}{(-i)e^{-i\pi 
kv_i}\theta\left[\hspace*{-6pt}\begin{array}{c}1/2+kv_i\\1/2\end{array}\hspace*{-6pt}\right](0,2il)\theta\left[\hspace*{-6pt}\begin{array}{c}kv_i\\0\end{array}\hspace*{-6pt}\right](0,2il)}\no\\&&\times\textrm{Tr 
} \gamma_{\Omega_k,3}^{-1}\gamma_{\Omega_k,3}^{T}.
\ea
For the Klein bottle we have the transformations
$t=2\tau_2,l=\frac{1}{t}$
\ba
Z_{\mathcal{K}}&=&
2^2\frac{(1-1)}{4N}\int\limits_0^\infty\!\!dl \sum_{k=0}^{N-1}
\frac{\theta\left[\hspace*{-6pt}\begin{array}{c}1/2\\0\end{array}\hspace*{-6pt}\right](0,il)}{\eta(il)^3}\no\\&&\qquad \times\prod_{i=1}^3 
\frac{\left|2\sin(2\pi kv_i)\right|}{4(\sin(\pi (kv_i+\frac{1}{2})))^2} 
\frac{\theta\left[\hspace*{-6pt}\begin{array}{c}1/2+2kv_i\\0\end{array}\hspace*{-6pt}\right](0,il)}{(-i)e^{-2i\pi 
kv_i} \theta\left[\hspace*{-6pt}\begin{array}{c}1/2+2kv_i\\1/2\end{array}\hspace*{-6pt}\right](0,il)}.\no\\&&
\ea
The ultraviolet contribution comes from $\tau_2\rightarrow 0$ or
$l\rightarrow \infty$. We have
\be
\lim_{l\rightarrow\infty}\frac{\theta\left[\hspace*{-6pt}\begin{array}{c}a\\b_1\end{array}\hspace*{-6pt}\right](0,il)}{\theta\left[\hspace*{-6pt}\begin{array}{c}a\\b_2\end{array}\hspace*{-6pt}\right](0,il)}=e^{2\pi 
ia(b_1-b_2)}.
\ee
In the sum $Z_{\mathcal{A}}+Z_{\mathcal{M}}+Z_{\mathcal{K}}$ the NS
and R contributions are both seperately free of ultraviolet
divergences, i.e. of tadpoles, under the condition that
\ba
0&=&\frac{1}{4}\prod_{i=1}^3 \left|2\sin(\pi kv_i)\right|\left(\textrm{Tr
} \gamma_{k,3}\right)^2 + 2\prod_{i=1}^3 
s_i(-2\sin(\pi kv_i)) \textrm{Tr
}(\gamma_{\Omega_k,3}^{-1}\gamma_{\Omega_k,3}^{T}) \no\\&&+4 
\prod_{i=1}^3 \frac{\left|2\sin(2\pi kv_i)\right|}{4(\sin(\pi
(kv_i+\frac{1}{2})))^2}.\label{tadpolecondition1D3a} 
\ea
We can choose $\textrm{Tr }(\gamma_{\Omega_k,3}^{-1}
\gamma_{\Omega_k,3}^{T})= \pm\textrm{Tr }\gamma_{2k,3}$,
where the positive sign is the $SO$ projection and the negative sign
is the $Sp$ projection. We have
\be
\frac{1}{4}\prod_{i=1}^3 \left|2\sin(\pi kv_i)\right|\left(\textrm{Tr
} \gamma_{k,3}\right)^2=\frac{1}{4}\prod_{i=1}^3 \left|2\sin(2\pi
kv_i)\right|\left(\textrm{Tr } \gamma_{2k,3}\right)^2.
\ee
Using $\sin(2\pi kv_i) = 2 \sin(\pi kv_i)\cos(\pi kv_i)$ and
$\sin(\pi(kv_i+\frac{1}{2})) = \cos(\pi kv_i)$
we find that (\ref{tadpolecondition1D3a}) is equivalent to
\be
0=\frac{1}{4}\left[\prod_{i=1}^3 \left|2\sin(2\pi kv_i)\right|\right]
\left(\textrm{Tr }\gamma_{2k,3}\mp 4\prod_{i=1}^3 \frac{1}{2\cos(\pi
kv_i)}\right)^2\label{tadpolecondition3D3a}
\ee
and is a perfect square.

\end{document}